\documentclass[apj]{emulateapj}
\usepackage{mathptmx}

\def\gtorder{\mathrel{\raise.3ex\hbox{$>$}\mkern-14mu
             \lower0.6ex\hbox{$\sim$}}}
\def\ltorder{\mathrel{\raise.3ex\hbox{$<$}\mkern-14mu
             \lower0.6ex\hbox{$\sim$}}}

\slugcomment{Submitted to ApJ}

\shorttitle{SN\,2010jl in a Massive Cocoon}

\shortauthors{Ofek et al.}

\begin{document}

\title{SN\,2010jl: Optical to hard X-ray observations reveal an explosion embedded in a ten solar mass cocoon}
\author{Eran O. Ofek\altaffilmark{1},
Andreas Zoglauer\altaffilmark{2}, 
Steven E. Boggs\altaffilmark{2},
Nicolas M. Barri\'{e}re\altaffilmark{2}, 
Stephen P. Reynolds\altaffilmark{3},
Chris L. Fryer\altaffilmark{4},
Fiona A. Harrison\altaffilmark{5},
S. Bradley Cenko\altaffilmark{6,7},
Shrinivas R. Kulkarni\altaffilmark{5},
Avishay Gal-Yam\altaffilmark{1},
Iair Arcavi\altaffilmark{1},
Eric Bellm\altaffilmark{5},
Joshua S. Bloom\altaffilmark{6},
Finn Christensen\altaffilmark{8},
William W. Craig\altaffilmark{9},
Wesley Even\altaffilmark{4},
Alexei V. Filippenko\altaffilmark{6},
Brian Grefenstette\altaffilmark{5},
Charles J. Hailey\altaffilmark{9},
Russ Laher\altaffilmark{10},
Kristin Madsen\altaffilmark{5},
Ehud Nakar\altaffilmark{11},
Peter E. Nugent\altaffilmark{12},
Daniel Stern\altaffilmark{13},
Mark Sullivan\altaffilmark{14},
Jason Surace\altaffilmark{10}, and
William W. Zhang\altaffilmark{7}
}
\altaffiltext{1}{Benoziyo Center for Astrophysics, Weizmann Institute
  of Science, 76100 Rehovot, Israel}
\altaffiltext{2}{Space Sciences Laboratory, Department of Physics, University of California, 7 Gauss Way, Berkeley, CA 94720, USA}
\altaffiltext{3}{Department of Physics, North Carolina State University, Raleigh, NC 27695-8202}
\altaffiltext{4}{CCS Division, Los Alamos National Laboratory, Los Alamos, NM 87545, USA}
\altaffiltext{5}{Cahill Center for Astronomy and Astrophysics, California Institute of Technology, Pasadena, CA 91125, USA}
\altaffiltext{6}{Department of Astronomy, University of California, Berkeley, CA 94720-3411, USA}
\altaffiltext{7}{Astrophysics Science Division, NASA Goddard Space Flight Center, Mail Code 661, Greenbelt, MD, 20771, USA}
\altaffiltext{8}{DTU Space-National Space Institute, Technical University of Denmark, Elektrovej 327, 2800 Lyngby, Denmark}
\altaffiltext{9}{Columbia Astrophysics Laboratory, 538 West 120th Street, New York, NY 10027, USA}
\altaffiltext{10}{Spitzer Science Center, MS 314-6, California Institute of Technology, Pasadena, CA 91125, USA}
\altaffiltext{11}{Raymond and Beverly Sackler School of Physics and Astronomy, Tel Aviv University, Tel Aviv 69978, Israel}
\altaffiltext{12}{Computational Cosmology Center, Lawrence Berkeley National Laboratory, 1 Cyclotron Road, Berkeley, CA 94720, USA}
\altaffiltext{13}{Jet Propulsion Laboratory, California Institute of Technology, Pasadena, CA 91109, USA}
\altaffiltext{14}{School of Physics and Astronomy, University of Southampton, Southampton SO17 1BJ, UK}

\begin{abstract}

Some supernovae (SNe) may be
powered by the interaction of the SN
ejecta with a large amount of circumstellar matter (CSM).
However, quantitative estimates of the CSM mass around such SNe
are missing when the CSM material is optically thick.
Specifically, current estimators are sensitive to uncertainties
regarding the CSM density profile and the ejecta velocity.
Here we outline a method to measure the mass of the optically thick CSM
around such SNe.
We present new visible-light and X-ray observations of SN\,2010jl (PTF\,10aaxf),
including the first detection of a SN in the hard X-ray band using {\it NuSTAR}.
The total radiated luminosity of SN\,2010jl is extreme, at least $9\times10^{50}$\,erg.
By modeling the visible-light data,
we robustly show that the mass of the circumstellar material
within $\sim 10^{16}$\,cm of the progenitor of SN\,2010jl was in excess of 10\,M$_{\odot}$.
This mass was likely ejected tens of years prior to the
SN explosion.
Our modeling suggests that the shock velocity during shock breakout
was $\sim6000$\,km\,s$^{-1}$, decelerating to $\sim2600$\,km\,s$^{-1}$
about two years after maximum light.
Furthermore, our late-time {\it NuSTAR} and {\it XMM} spectra of the SN
presumably provide the first direct measurement of SN shock velocity
two years after the SN maximum light --- measured to be in the range
of 2000\,km\,s$^{-1}$ to 4500\,km\,s$^{-1}$
if the ions and electrons are in equilibrium, and $\gtorder2000$\,km\,s$^{-1}$
if they are not in equilibrium.
This measurement is in agreement with the shock velocity
predicted by our modeling of the visible-light data.
Our observations also show that the average radial density distribution of the
CSM roughly follows an $r^{-2}$ law.
A possible explanation for the $\gtorder10$\,M$_{\odot}$ of CSM
and the wind-like profile is that they are the result of
multiple pulsational pair instability events prior to the SN explosion,
separated from each other by years.

\end{abstract}


\section{Introduction}
\label{sec:Introduction}

Some supernovae (SNe),
especially of Type IIn (for a review, see Filippenko 1997),
show strong evidence
for the existence of a large amount
(i.e., $\gtorder10^{-3}$\,M$_{\odot}$)
of circumstellar matter (CSM)
ejected months to years prior to the SN explosion
(e.g., Dopita et al. 1984;
Weiler et al. 1991;
Chugai \& Danziger 1994;
Chugai et al. 2003;
Gal-Yam et al. 2007;
Gal-Yam \& Leonard 2009;
Ofek et al. 2007, 2010, 2013b;
Smith et al. 2007, 2008, 2009;
Kiewe et al. 2012).
In some cases even larger CSM masses,
of order 10\,M$_{\odot}$, have been reported.
However, these claims are based on very rough modeling that
may suffer from more than an order of magnitude uncertainty
(e.g., see Moriya \& Tominaga 2012 for discussion).
Interestingly, three SNe were recently reported
to show outbursts taking place prior to the SN explosion
(e.g., Pastorello et al. 2007, 2013; Foley et al. 2007, 2011;
Mauerhan et al. 2012; Ofek et al. 2013b).

Interaction of the SN blast wave with the CSM
in many cases produces long-lived panchromatic signals
from radio to X-ray energies
(e.g., Slysh 1990;
Chevalier \& Fransson 1994;
Chevalier 1998; Weiler et al. 1991;
Chandra et al. 2012a, 2012b; Ofek et al. 2013a).
Most important for the interpretation
of the light curves of some SNe~IIn,
Svirski, Nakar, \& Sari (2012) have
presented predictions for the optical and X-ray
luminosity evolution of SNe powered by
interaction of their ejecta with the CSM.
Observing these signals has
the potential to both unveil the physical parameters
of the explosion and to measure the CSM mass.

Until recently, hard X-ray instruments lacked
the sensitivity to study SN shock interactions.
However, with the launch of the
Nuclear Spectroscopic Telescope Array ({\it NuSTAR})
focusing hard X-ray space telescope (Harrison et al. 2013),
it is now possible to measure
the hard X-ray spectrum (3--79\,keV) of such events.
This in turn has the potential to directly measure, in some cases,
the shock velocity of the SN,
which is hard to estimate using other proxies.
Here we present the first detection of a supernova (SN\,2010jl,
also known as PTF\,10aaxf)
outside the Local Group
in the hard X-ray band using {\it NuSTAR}.

SN\,2010jl was discovered on 2010 Nov. 3.5
(Newton \& Puckett 2010) in the star-forming galaxy UGC\,5189A
(redshift $z=0.0107$, distance 49\,Mpc),
and was classified as a SN~IIn
(Benetti et al. 2010; Yamanaka et al. 2010).
The SN coordinates, as measured in images taken by the
Palomar Transient Factory, are $\alpha=$09$^{\rm h}$42$^{\rm m}$53.$^{\rm s}$337,
$\delta=+$09$^{\circ}$29$^{'}$42$^{''}.13$ (J2000.0).
Pre-discovery images suggest
that the SN exploded prior to 2010 Sep. 10 (Stoll et al. 2011).
However, the rise time and explosion date are not well constrained.
Based on analysis of archival {\it Hubble Space Telescope}
images, Smith et al. (2010) argued that the progenitor mass
is $\gtorder 30$\,M$_{\odot}$.
Stoll et al. (2011) show that the SN host galaxy has
a metallicity of $\ltorder 0.3$ solar.

Patat et al. (2011) report on spectropolarimetry of SN\,2010jl
obtained about 15\,days after its discovery.
They find a significant, and almost constant with wavelength, linear
polarization level (1.7\%--2.0\%) with constant
position angle. Based on that, they suggest that the
axial ratio of the photosphere of the event is $\ltorder0.7$.
They also note that the Balmer-line cores have small
polarization, indicating that they form above
the photosphere. They also argue that at the epoch of their
observations, the CSM had a very low dust content.

Soon after its discovery,
SN\,2010jl was detected in X-rays
(Chandra et al. 2012a; Ofek et al. 2013a).
Chandra et al. (2012a) analyzed the first two {\it Chandra}
observations of this source.
They find a high bound-free absorption column density,
roughly $10^{24}$\,cm$^{-2}$,
about one month after SN maximum light,
decreasing to $\sim3\times10^{23}$\,cm$^{-2}$ about one
year after maximum light.
However, the value of the column density depends on the assumed emission model.
Chandra et al. (2012a)
reported that the hardest X-ray component in the SN\,2010jl spectra
has a temperature
above 8\,keV, but given the {\it Chandra} drop in sensitivity
above 8\,keV, this temperature is not well
constrained.
Here we also reanalyze the {\it Chandra} observations.
Based on the X-ray observations of SN\,2010jl,
Ofek et al. (2013a) suggested that the optical luminosity of this
SN is powered by shock breakout in an optically thick CSM.

Here we analyze {\it NuSTAR}, {\it XMM-Newton}, {\it Chandra}, and {\it Swift}-XRT
as well as visible-light and ultraviolet (UV) observations
of the extraordinary Type IIn SN\,2010jl.
Under the conditions we show to hold for it,
at early times after explosion the shock in
the dense wind is radiation dominated. That is, the energy
density behind the shock is primarily in radiation because of the high
Thomson optical depths. In this
case, the shock breaks out (i.e., is detectable to a distant observer) when
the photon diffusion time is comparable to the dynamical time. 
Straightforward considerations relate the shock radius,
velocity, mass in the wind ahead of the shock, and luminosity, so that
the CSM mass can be inferred.  We generalize earlier discussions
to different power-law profiles for the wind and the SN ejecta to obtain
general relations among these quantities, and apply them to optical and
X-ray observations of SN\,2010jl.
Combining our model with the observations,
we are able to measure the total CSM mass,
its density profile, and the temporal evolution of the shock velocity.

We note that throughout the paper dates
are given in the UTC system, and unless specified
differently errors represent the 1-$\sigma$ uncertainties.
The structure of this paper is as follows.
We present the observations in \S\ref{sec:Observations},
and the reduction of the X-ray data is discussed in \S\ref{sec:xspec}.
Our model is described in \S\ref{sec:model}.
In \S\ref{sec:obsmodel} we apply the model to the observations,
and we discuss our results in \S\ref{sec:Disc}.

\section{Observations}
\label{sec:Observations}

We obtained multi-wavelength observations of SN\,2010jl.
The most constraining observations for our model are
the bolometric light curve of the SN,
and the late-time X-ray spectrum obtained by {\it NuSTAR}$+${\it XMM}.
We note that the bolometric light curve is derived from the $R$-band observations
with a bolometric correction that we estimate from
the {\it Swift}-UVOT and spectroscopic observations.

\subsection{Visible light observations}

\begin{figure*}
\centerline{\includegraphics[width=16cm]{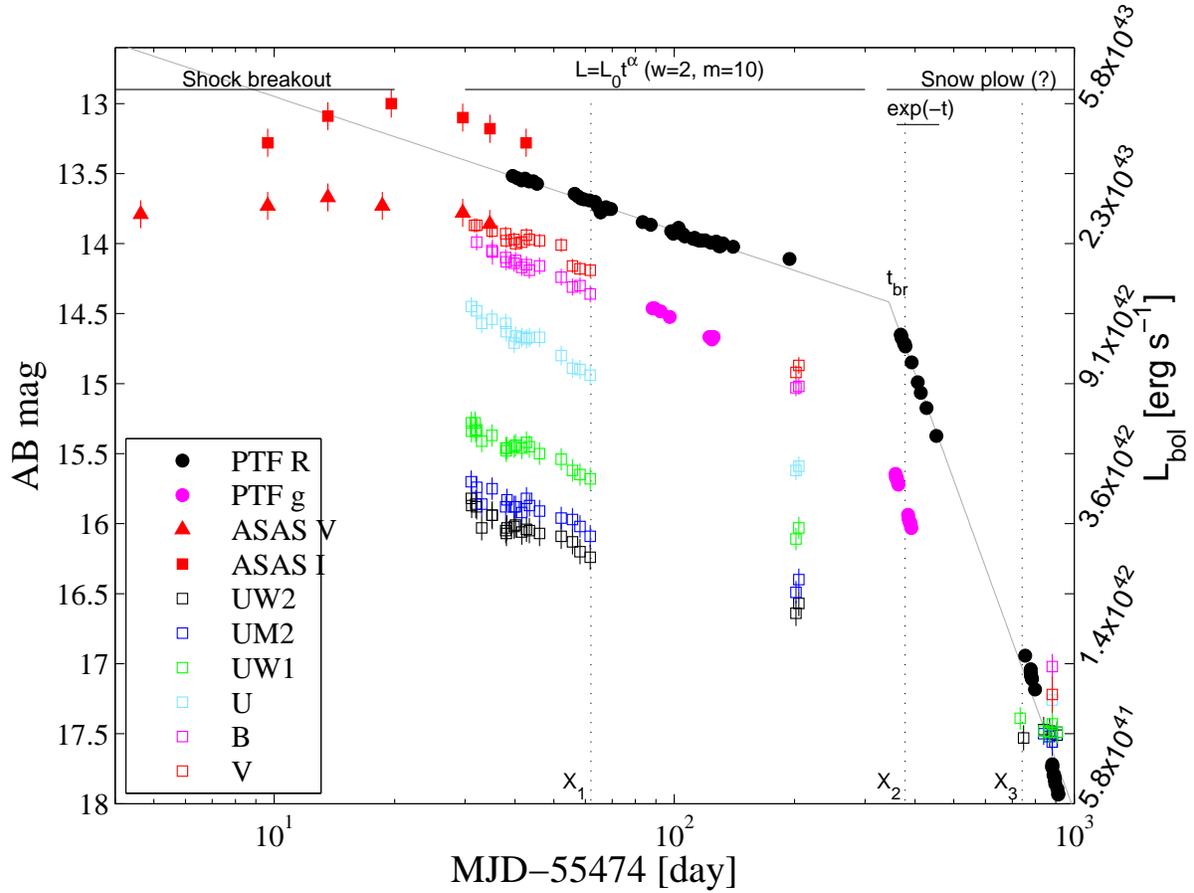}}
\caption{Optical light curves of SN\,2010jl.
The black filled cirles and magenta filled circles represent the PTF measurements, which
are based on image subtraction.
In this case the uncertainties include the Poisson error and a 0.015\,mag systematic
error added in quadrature (Ofek et al. 2012a, 2012b).
See the legend for ASAS and {\it Swift}-UVOT measurements.
The gray lines show the best-fit broken power law to the PTF $R$-band data.
The power-law index before (after) the break is $-0.38$ ($-3.14$).
The power-law break is at day 344 (with respect to MJD 55474).
The epochs of the {\it Chandra} and {\it NuSTAR}$+${\it XMM}
observations are marked by vertical dotted lines.
The right-hand ordinate axis shows the bolometric luminosity for the PTF $R$-band data,
assuming the bolometric correction is $-0.27$\,mag.
Time is measured from 20\,days prior to $I$-band maximum light.
The various physical stages are indicated at the top of the plot.
These are the shock breakout phase,
the early power-law decay,
and the snow-plow phase (see \S\ref{sec:model}).
Also shown is the section of the light curve which is fitted well by an exponential decay (i.e., ``$\exp(-t)$'').}
\label{fig:PTF10aaxf_LC}
\end{figure*}

The Palomar Transient Factory
(PTF\footnote{http://www.astro.caltech.edu/ptf/};
Law et al. 2009; Rau et al. 2009) detected SN\,2010jl
(PTF\,10aaxf) on 2010 Nov. 13.4,
ten days after its discovery by Newton \& Puckett (2010).
The PTF data-reduction pipeline is presented by Laher et al. (in prep.),
and the photometric calibration is described by Ofek et al. (2012a, 2012b).
The PTF light curve of this SN as well
as the All Sky Automated Survey (ASAS) prediscovery data points from Stoll et al. (2011)
are presented in Figure~\ref{fig:PTF10aaxf_LC}
and listed in Table~\ref{tab:PhotObs}.
ASAS first detected the SN on 2010 Sep. 10,
about 15\,days prior to $I$-band maximum light --- soon after
its solar conjunction.

\begin{deluxetable}{lllll}
\tablecolumns{5}
\tablewidth{0pt}
\tablecaption{Photometric observations}
\tablehead{
\colhead{Telescope}     &
\colhead{Filter}        &
\colhead{MJD-55474}     &
\colhead{Mag}           &
\colhead{Err}           \\
\colhead{}              &
\colhead{}              &
\colhead{day}           &
\colhead{mag}            &
\colhead{mag}            
}
\startdata
PTF        & R       & -178.762  &  $<21.8$ & \nodata \\
PTF        & R       &   39.444  &  13.514  &  0.003  \\
PTF        & R       &   39.487  &  13.519  &  0.002  \\
PTF        & R       &   40.489  &  13.532  &  0.004  \\
PTF        & R       &   40.533  &  13.532  &  0.004 
\enddata
\tablecomments{PTF, ASAS (Stoll et al. 2011), and {\it Swift}-UVOT
photometric observations of SN\,2010jl.
Time is measured relative to MJD 55474 (20 days prior to the $I$-band peak magnitude).
The PTF and {\it Swift} magnitudes are given in the AB system, while the
ASAS measurements are in the Vega system.
This table is published in its entirety in the electronic edition of
{\it ApJ}. A portion of the full table is shown here for
guidance regarding its form and content.}
\label{tab:PhotObs}
\end{deluxetable}

The first-year PTF flux measurements taken before MJD 55760
show a clear power-law decline (\S\ref{sec:obsmodel}),
the second-year flux measurements obtained between MJD 55760 and MJD 56070
(Fig.~\ref{fig:PTF10aaxf_LC})
are consistent with an exponential decay
(i.e., $\propto\exp(-t/\tau_{{\rm exp}})$, where $t$ is the time
and $\tau_{{\rm exp}}$ is the exponential time scale).
We find that the best-fit exponential time scale
is $\tau_{{\rm exp}}=129.8\pm1.5$\,day
($\chi^{2}/{\rm dof} = 0.7/15$),
where the uncertainty is estimated using the bootstrap technique (Efron 1982; Efron \& Tibshirani 1993).
We note that this is longer than the time scale expected from
$^{56}$Co decay ($\sim 111$\,day).
Were this decay produced by $^{56}$Ni decay to $^{56}$Co and
finally $^{56}$Fe, then
at least 27\,M$_{\odot}$ of $^{56}$Ni would be required, which
is unlikely.
Moreover, at later times the decay rate becomes significantly
slower than the exponential decay expected from radioactive material (see Fig.~\ref{fig:PTF10aaxf_LC}).
Therefore, a more reasonable interpretation is
that the SN light curve is powered by interaction of the SN shock with CSM.
Interestingly, the second-year and third-year data (MJD $>56070$)
are also roughly consistent with a power-law decay.
The power-law fits to the light-curve data are shown in Figure~\ref{fig:PTF10aaxf_LC}
and discussed in \S\ref{sec:obsmodel}.

\subsection{Spectroscopy}
\label{sec:spec}

\begin{figure}
\centerline{\includegraphics[width=8.5cm]{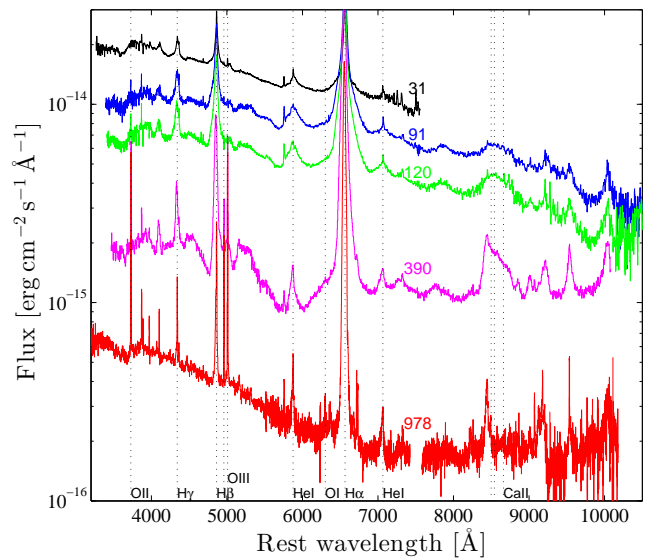}}
\caption{Selected visible-light spectra of SN\,2010jl. The number near each spectrum marks
its age in days (see Table~\ref{tab:LogSpec}).
The last spectrum taken on day 978 may be contaminated by emission from
the underlying star-forming region.
\label{fig:SN2010jl_Spec}}
\end{figure}

SN\,2010jl was observed spectroscopically by the PTF collaboration
on several occasions. A log file of the observations is presented
in Table~\ref{tab:LogSpec}.
The data will be electronically released via
the WISeREP website\footnote{http://www.weizmann.ac.il/astrophysics/wiserep/}
(Yaron \& Gal-Yam 2012).
Selected spectra of SN\,2010jl are shown in Figure~\ref{fig:SN2010jl_Spec}.
\begin{deluxetable}{lllll}
\tablecolumns{5}
\tablewidth{0pt}
\tablecaption{Visible-light spectroscopic observations}
\tablehead{
\colhead{MJD}           &
\colhead{Day}           &
\colhead{Telescope}     &
\colhead{Instrument}    &
\colhead{$T_{{\rm eff}}$} \\
\colhead{day}         &
\colhead{day}         &
\colhead{}            &
\colhead{}            &
\colhead{K}           
}
\startdata
 55505 &  31 & Keck     & LRIS   &     7560       \\
 55507 &  33 & Keck     & DEIMOS &     6800       \\
 55507 &  33 & Keck     & DEIMOS &     6320       \\
 55515 &  41 & Lick     & Kast   &     7090       \\
 55530 &  56 & Lick     & Kast   &     7360       \\
 55538 &  64 & Keck     & LRIS   &     7160       \\
 55565 &  91 & Lick     & Kast   &     6590       \\
 55587 & 113 & Lick     & Kast   &     6650       \\
 55594 & 120 & Lick     & Kast   &     6740       \\
 55864 & 390 & P200     & DBSP   &     6380       \\
 56332 & 858 & Keck     & LRIS   &    10400       \\
 56414 & 940 & P200     & DBSP   &    10600       \\
 56421 & 947 & Keck     & LRIS   &    11670       \\
 56452 & 978 & Keck     & LRIS   &     9350
\enddata
\tablecomments{MJD is the modified Julian day.
Day is the time relative to MJD 55474 (i.e., 20\,days before the $I$-band peak flux). The formal uncertainties in the temperature measurements are about 50--300\,K.
However, due to metal-line blanketing, the actual effective temperature can be
higher.
A large fraction of the spectroscopic observations listed here were
presented and discussed in Smith et al. (2012).
\label{tab:LogSpec}}
\end{deluxetable}

\begin{figure}
\centerline{\includegraphics[width=8.5cm]{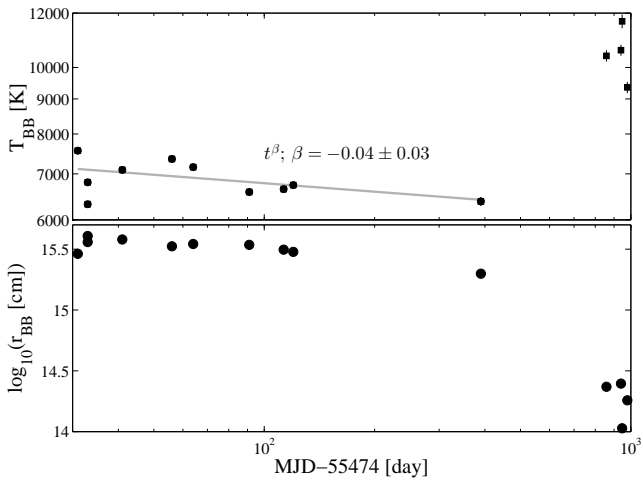}}
\caption{Temperature and radius of a black body that best fits the visible-light spectroscopic
observations as a function of time. Before fitting the spectra,
we corrected the flux normalization by comparing the spectra
synthetic photometry with the PTF $R$-band magnitudes.
We also removed the prominent emission lines and the Balmer
discontinuity. We note that because of additional metal-line blanketing, this estimate is likely
a lower limit on the actual temperature.
The gray line shows the best-fit power law
to the temperature measurements in the first 390\,days.
The measurements marked by squares were
obtained clearly after the break in the optical light curve and were not used
in the fit of the temperature as a function of time.
These late-time measurements may be contaminated by
the host-galaxy light.
\label{fig:SN2010jl_TempRad_Spec}}
\end{figure}

Inspection of the spectra of SN\,2010jl show that the H$\alpha$ line
consists of several components.
The narrowest features we detected are H$\alpha$, H$\beta$, and He\,I P-Cygni lines,
with a velocity difference between the peak and minimum of $\sim 70$\,km\,s$^{-1}$
(see also Smith et al. 2012).
The H$\alpha$ profile in the spectra
can be decomposed into a Lorentzian and a Gaussian, where the Gaussian has a velocity width of
$\sigma \approx 300$\,km\,s$^{-1}$. 
Alternatively, the early-time spectra can be decomposed into three Gaussians,
in which the widest Gaussian has velocity width $\sigma \approx 4000$\,km\,s$^{-1}$.
At late times, about six months after maximum light, the H$\alpha$ line
develops some asymmetry; it is discussed by Smith et al. (2012)
and attributed to dust formation.
We fitted a black-body spectrum to the spectroscopic measurements as a function of time,
and the derived black-body temperatures and radii are shown in Figure~\ref{fig:SN2010jl_TempRad_Spec}.

\subsection{Swift-UVOT}
\label{sec:UVOT}

\begin{figure}
\centerline{\includegraphics[width=8.5cm]{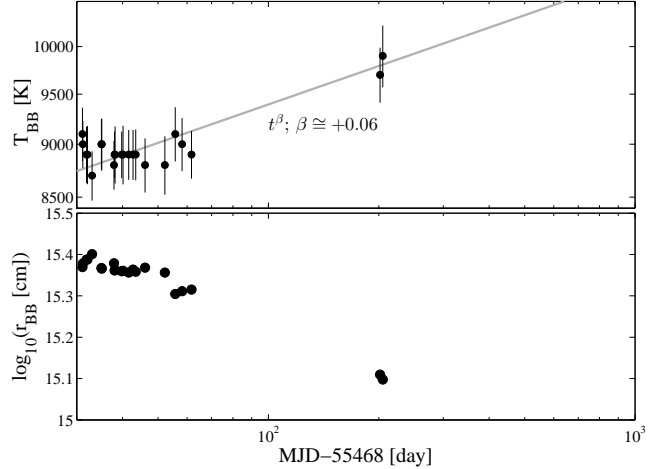}}
\caption{Temperature and radius of a black body that best fits the {\it Swift}-UVOT
observations as a function of time. Observations made more than 500\,days
after maximum light are excluded, as they are significantly affected by the host-galaxy light
and we do~not yet have a reference image of the host.
The gray line shows a power law fitted to the temperature data.
}
\label{fig:SN2010jl_TempRad_UVOT}
\end{figure}

The Ultra-Violet/Optical Telescope (UVOT; Roming et al. 2005)
onboard the {\it Swift} satellite (Gehrels et al. 2004) observed SN\,2010jl on several occasions.
The data were reduced using standard procedures (e.g., Brown et al. 2009).
Flux from the transient was extracted from a $3''$-radius aperture, with a correction applied
to put the photometry on the standard UVOT system (Poole et al. 2008).
The resulting measurements, all of which have been converted to the AB system, are listed in
Table~\ref{tab:PhotObs} and are shown in Figure~\ref{fig:PTF10aaxf_LC}.
We caution that these results have not incorporated any contribution from the underlying host galaxy,
and may therefore overestimate the SN flux at late times.
Specifically, the UVOT measurements in Figure~\ref{fig:PTF10aaxf_LC} near 900\,days
are heavily contaminated by an underlying star-forming region in the host galaxy.
We fitted a black-body spectrum to the UVOT measurements as a function of time,
and the results are shown in Figure~\ref{fig:SN2010jl_TempRad_UVOT}.
In the fits we corrected the flux measurements for Galactic extinction,
assuming $E_{B-V}=0.027$\,mag (Schlegel et al. 1998) and $R_{V}=3.08$ (Cardelli et al. 1989).
We note that we also tried to fit the black-body spectrum with
$E_{B-V}$ as a free parameter, and verified that the best fit is
obtained near the Schlegel et al. (1998) value for $E_{B-V}$.
The {\it Swift}-derived black-body temperature shows some indications
that it is rising in the first $\sim200$\,days after maximum light.
However, we caution that deviations from a black body caused by spectral lines
that are not dealt with in the broad-band observations,
as well deviations from a black-body spectrum (see \S\ref{sec:ModelLate})
and metal-line blanketing, can affect the derived temperature and radius.
Therefore, we argue that the quoted temperatures
are likely only a lower limit on the effective temperatures.

These temperature measurements differ from those obtained using the spectroscopic observations
(\S\ref{sec:UVOT}).
However, due to metal-line blanketing and given that the spectral peak is too blue to be probed
by visible-light spectra, we consider both the spectroscopic
and UVOT observations to be lower limits on the temperature.
The temperature evolution based on the visible-light spectra is opposite
to that based on the UVOT observations.
However, both evolutions seen in figures
\ref{fig:SN2010jl_TempRad_Spec}
and \ref{fig:SN2010jl_TempRad_UVOT} are very moderate.
In \S\ref{sec:ModelEarly} we investigate the effect of this uncertainty
on our results,
and in \S\ref{sec:ModelLate} we discuss
the nature of the decrease in the black-body radius at late times.

\subsection{NuSTAR}
\label{sec:NuStar}

{\it NuSTAR} is the first hard X-ray focusing satellite (Harrison et al. 2013).
Its broad energy range (3--79\,keV) allows us to determine the previously unconstrained temperature of the hardest component of the X-ray spectrum.
{\it NuSTAR} observed SN\,2010jl on 2012 Oct. 6, roughly 2\,yr after the discovery of the SN.
We obtained a usable exposure time of 46\,ks. This was the first SN observed by the {\it NuSTAR} ``supernovae and
target-of-opportunity program.'' Spectra and images where extracted using the standard {\it NuSTAR} Data Analysis Software
({\tt NuSTARDAS} version 0.11.1) and {\tt HEASOFT} (version 6.13). {\tt XSPEC} (Arnaud 1996, version 12.8)
was used to perform the spectral analysis in combination with the {\it XMM} data.
A summary of the high-energy observations
of SN\,2010jl is given in Table~\ref{tab:LogXray}.

\subsection{XMM}
\label{sec:XMM}

Shortly after we obtained the {\it NuSTAR} observations, we triggered {\it XMM-Newton} for a target-of-opportunity
observation (see Table~\ref{tab:LogXray}) with the goal of determining the bound-free
absorption utilizing {\it XMM}'s good low-energy X-ray response.
The observation was carried out during 2012 Nov. 1 for 13\,ks, resulting in a usable exposure time of $\sim 10$\,ks
for the MOS1 and MOS2 detectors
and $\sim 4$\,ks for the PN detector, after filtering out periods of high background flaring activity.
The Science Analysis System software ({\tt SAS}, version 12) was used for data reduction.
Spectral analysis combined with the {\it NuSTAR} data was performed using {\tt XSPEC} version 12.8.

\begin{deluxetable}{lllll}
\tablecolumns{5}
\tablewidth{0pt}
\tablecaption{Log of {\it NuSTAR}, {\it XMM}, and {\it Chandra} Observations}
\tablehead{
\colhead{Inst}            &
\colhead{ObsID}           &
\colhead{MJD}             &
\colhead{Exposure time}   &
\colhead{Count rate}      \\
\colhead{}        &
\colhead{}        &
\colhead{(day)}     &
\colhead{(s)}      &
\colhead{(ct\,ks$^{-1}$)} 
}
\startdata
{\it Chandra}  & 11237      & 55522.12 & 10046 & $11.2\pm1.0$     \\
{\it Chandra}  & 11122      & 55537.29 & 19026 & $9.64\pm0.71$    \\
{\it Chandra}  & 13199      & 55538.16 & 21032 & $11.72\pm0.74$   \\
{\it Chandra}  & 13781      & 55852.09 & 41020 & $32.05\pm0.88$   \\
{\it NuSTAR}   & 40002092001& 56205.98 & 46000 & $27.2\pm0.8$     \\
{\it XMM}      & 0700381901 & 56232.72 & 12914 & $158\pm4$        
\enddata
\tablecomments{MJD is the modified Julian day.
The background-corrected count rate is in the 0.2--10\,keV
band for {\it Chandra} and {\it XMM}, and 3--79\,keV
for {\it NuSTAR}.
For {\it Chandra} we used an extraction aperture radius of $3''$ and a sky annulus
whose inner (outer) radius is $20''$ ($40''$).
For {\it XMM} we used an extraction aperture radius of $32''$ and a sky annulus
whose inner (outer) radius is $32''$ ($33.5''$).
For {\it NuSTAR} we used an extraction aperture radius of $60''$ and a sky annulus
whose inner (outer) radius is $60''$ ($100''$).
The {\it XMM} count rate is the combined value from all three instruments.
The first three {\it Chandra} observations were obtained within a time window of 16\,days.
Here we analyzed the first three observations jointly, and refer to them
as the first {\it Chandra} epoch,
while {\it Chandra} ObsID 13781 is referred to as the {\it Chandra} second epoch.}
\label{tab:LogXray}
\end{deluxetable}

\subsection{Chandra}
\label{sec:Chandra}

{\it Chandra} observed the location of SN\,2010jl on five epochs
(PIs Pooley, Chandra; Chandra et al. 2012a).
All the observations except one are public.

\begin{figure}
\centerline{\includegraphics[width=8.5cm]{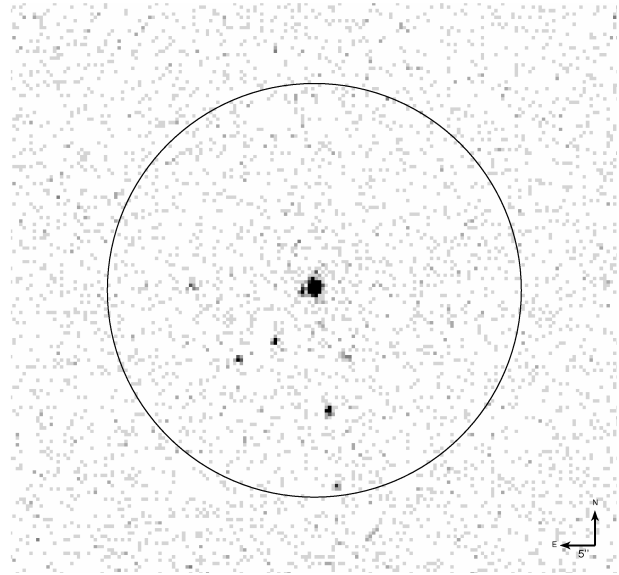}}
\caption{{\it Chandra} image of SN\,2010jl (ObsID 13781). The SN is the bright source
at the center. The nearby source is $2''$ east of the SN. The black circle has a radius
of $30''$, similar to the {\it XMM} extraction region.
Several sources are visible within this extraction radius.
There are no additional bright sources outside this radius and within $60''$
of the SN position (i.e., the {\it NuSTAR} extraction region).}
\label{fig:SN2010jl_Chandra13781}
\end{figure}

Inspection of the {\it Chandra} images shows
emission from the SN position,
as well as from another source only about $2''$ east of the
SN (Fig.~\ref{fig:SN2010jl_Chandra13781}). 
In order to make sure that the {\it Chandra} flux measurements
are compatible with the other X-ray observations
we used a relatively large aperture of radius $3''$.
This extraction aperture
contains light from the nearby source.
The background was extracted from an annulus with an inner (outer)
radius of 20$''$ (40$''$). 
The observations are plotted in Figure~\ref{fig:SN2010jl_XLC_MeanE}
and presented in Table~\ref{tab:LogXray}.

In addition, there are multiple weak sources located within the source extraction regions of
{\it XMM} and {\it NuSTAR} (Figure~\ref{fig:SN2010jl_Chandra13781}). We use
the {\it Chandra} observation to determine their mean flux
and spectrum, and as an additional (known) component while fitting the spectra
from {\it NuSTAR} and {\it XMM}.
The {\it Chandra} data were analyzed using
{\tt XSPEC}\footnote{http://heasarc.gsfc.nasa.gov/xanadu/xspec/}
V12.7.1 (Schafer 1991).
The Galactic neutral hydrogen column density in the direction of SN\,2010jl is
$N_{{\rm H}}=3\times10^{20}$\,cm$^{-2}$ (Dickey \& Lockman 1990)
All of the nearby sources were fitted jointly
with an absorbed power law assuming Galactic absorption.
The fit resulted in a photon power-law index of $\Gamma=1.375$ and a flux of $6.3\times10^{-6}$\,ph\,cm$^{-2}$\,s$^{-1}$
in the energy range 0.3--10\,keV
($\chi^{2}$/dof $=12.5/12$).
The spectra as well as the contamination by the nearby sources
are discussed and modeled in \S\ref{sec:xspec}.

\subsection{Swift-XRT}
\label{sec:XRT}

\begin{figure}
\centerline{\includegraphics[width=8.5cm]{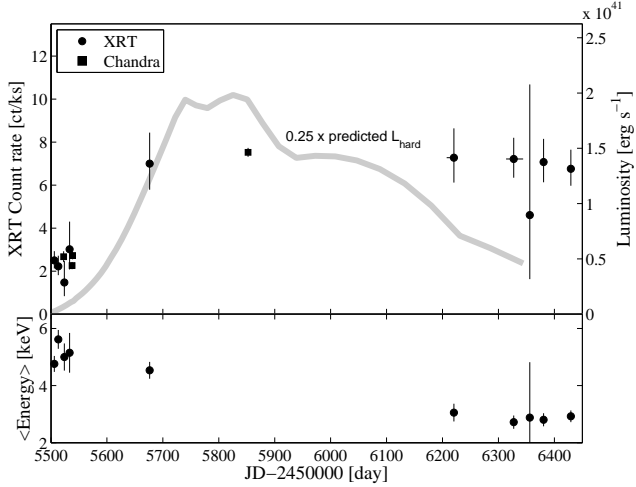}}
\caption{{\it Upper panel:} X-ray light curve of SN\,2010jl.
The left-hand ordinate axis shows the {\it Swift}-XRT count rate, while the right-hand ordinate axis
represents the {\it Swift} and {\it Chandra} X-ray luminosity in the 0.2--10\,keV
band assuming a Galactic neutral hydrogen column density of
$3\times10^{20}$\,cm$^{-2}$ (Dickey \& Lockman 1990) and an X-ray spectrum of the form
$n(E)\propto E^{-1}$, where $n(E)$ is the number of photons per unit energy.
We note that the unabsorbed luminosity would be a factor of
1.7, 4.5, and 54 times higher for a neutral hydrogen column
density of $10^{22}$, $10^{23}$, and $10^{24}$\,cm$^{-2}$, respectively.
The gray line shows $0.25$ times the predicted X-ray luminosity based on
Equation~\ref{eq:Lx}, assuming $m=10$ before $t_{{\rm br}}$
and $m=4$ afterward (see \S\ref{sec:ModelX}).
The XRT and {\it Chandra} measurements are contaminated by
the nearby source and therefore over estimate the flux by about 10\%.
{\it Lower panel:} The mean X-ray energy of the {\it Swift}-XRT photons
in the 0.2--10\,keV range.}
\label{fig:SN2010jl_XLC_MeanE}
\end{figure}

The {\it Swift} X-Ray Telescope (XRT; Burrows et al. 2005)
observed SN\,2010jl on multiple epochs since the SN discovery.
For each {\em Swift}/XRT image of the SN,
we extracted the number of X-ray counts in the 0.2--10\,keV
band within an aperture of $9''$ radius
centered on the SN position.
This aperture contains $\sim 50$\% of
the source flux (Moretti et al.\ 2004).
The background count rates were estimated in
an annulus around the SN location, with an inner (outer) radius of $50''$ ($100''$).
The log of {\it Swift}-XRT observations,
along with the source and background X-ray counts in the individual
observations, are listed in Table~\ref{tab:XRTobs}.
The binned {\it Swift}-XRT observations are presented in
Figure~\ref{fig:SN2010jl_XLC_MeanE} and listed in Table~\ref{tab:XRTbin}.
\begin{deluxetable}{llll}
\tablecolumns{4}
\tablewidth{0pt}
\tablecaption{{\it Swift}-XRT observations}
\tablehead{
\colhead{MJD}          &
\colhead{Exposure time}          &
\colhead{Source}      &
\colhead{Background}     \\
\colhead{day} &
\colhead{(ks)} &
\colhead{(ct)}            &
\colhead{(ct)}
}
\startdata
55505.08&  1.93&   3&   2\\
55505.15& 13.49&  15&  26\\
55505.67&  6.70&  12&  19\\
55505.89&  4.66&   6&  19\\
55506.08&  1.80&   0&   2
\enddata
\tablecomments{MJD is the modified Julian day.
Source is the number of counts in the 0.2--10\,keV band
within an aperture radius of $9''$, centered on the source position.
Background is the number of counts
in the 0.2--10\,keV band
in an annulus of inner (outer)
radius $50''$ ($100''$) around the source.
The ratio between the background annulus area and the aperture area
is $92.59$.
This table is published in its entirety in the electronic edition of
{\it ApJ}. A portion of the full table is shown here for
guidance regarding its form and content.}
\label{tab:XRTobs}
\end{deluxetable}
\begin{deluxetable}{llllll}
\tablecolumns{6}
\tablewidth{0pt}
\tablecaption{Binned {\it Swift}-XRT data}
\tablehead{
\colhead{$\langle{\rm MJD}\rangle$} &
\multicolumn{2}{c}{Range}           &
\colhead{CR}                        &
\colhead{Exp.}                      &
\colhead{$\langle E \rangle$}      \\
\colhead{(day)}                 &
\colhead{(day)}                 &
\colhead{(day)}                 &
\colhead{(cnt/ks)}              &
\colhead{(ks)}                  &
\colhead{(keV)}            
}
\startdata
 55505.5&$-0.4$ & 3.6 & $2.50_{-0.37}^{+0.43}$  &  36.02 &   $4.76\pm0.28$\\
 55512.6&$-0.7$ & 4.9 & $2.22_{-0.41}^{+0.49}$  &  26.08 &   $5.62\pm0.33$\\
 55523.4&$-3.3$ & 2.8 & $1.47_{-0.64}^{+1.0}$   &   6.79 &   $5.00\pm0.48$\\
 55532.7&$-3.0$ & 3.0 & $3.02_{-0.94}^{+1.3}$   &   6.63 &   $5.15\pm0.70$\\
 55675.9&$-0.3$ & 3.2 & $7.00_{-1.2}^{+1.4}$    &   9.43 &   $4.53\pm0.30$\\
 56219.9&$-12.9$& 1.4 & $7.28_{-1.2}^{+1.4}$    &  10.72 &   $3.06\pm0.31$\\
 56326.9&$-13.8$&16.7 & $7.21_{-0.87}^{+1.0}$   &  18.85 &   $2.72\pm0.23$\\
 56355.5&$-0.0$ & 0.0 & $4.61_{-3.0}^{+6.1}$    &   0.87 &   $2.88\pm1.94$\\
 56380.3&$-0.0$ & 0.0 & $7.07_{-0.94}^{+1.1}$   &  15.83 &   $2.80\pm0.23$\\
 56429.1&$-2.4$ & 4.4 & $6.76_{-0.79}^{+0.89}$  &  21.60 &   $2.92\pm0.21$
\enddata
\tablecomments{SN\,2010jl binned {\it Swift}-XRT light curve.
$\langle{\rm MJD}\rangle$ is the weighted mean modified Julian day
of all the observations in a given bin,
where the observations are weighted by their exposure times.
Range is the time range around $\langle{\rm MJD}\rangle$
in which the light curve
(Table~\ref{tab:XRTobs}) was binned.
CR is the counts rate along
with the lower and upper 1$\sigma$ uncertainties.
The source count rates are corrected
for extraction aperture losses (50\%).
$\langle E \rangle$ is the mean energy of photons within the 0.2--10\,keV range
and the standard error of the mean.
}
\label{tab:XRTbin}
\end{deluxetable}

\section{X-ray Spectra of SN\,2010jl}
\label{sec:xspec}

Chandra et al. (2012a) analyzed the {\it Chandra} spectra.
They found that multiple components are required
(e.g., two Mekal\footnote{Mekal is an emission spectrum from hot diffuse gas
with lines from Fe, as well as several other elements.}
spectra; Mewe et al. 1986) in order to obtain
a good fit.
Based on our modeling described in \S\ref{sec:obsmodel},
we argue that this SN is powered by interaction of the SN shock
with an optically thick CSM.
In this case, at least in the first 2\,yr after discovery, using Mekal
(i.e., optically thin emission) components is not physically justified.
It is possible that the good fit
obtained by Chandra et al. (2012)
is a result of the large number of free parameters in their model.
In addition, it is possible that the low-energy component
suggested by Chandra et al. (2012) originates
from the nearby (soft) source (see below).
Here we attempt to fit physically motivated
simple models, with a small number of degrees of freedom. 

As mentioned in \S\ref{sec:Chandra}, the {\it Chandra} images show
several other sources near the position of SN\,2010jl.
Interestingly, we identify one source
only $2''$ from the SN position.
We note that the mean photon energies of the primary source (i.e., the SN)
and this nearby source are very different, about 4 and 2\,keV, respectively.
We fitted a two point-spread function (PSF; {\tt CALDB}, version 4.5.5.1) model
to the two sources simultaneously using our own code.
We use the {\it Chandra} 4\,keV PSF for the SN, and the 2\,keV PSF for the
nearby source. This exercise allows us to measure the flux
of the nearby source (which is useful as a constraint while analyzing data from
other instruments with poorer resolution).
This also shows that the nearby source is real and not an artifact
of the {\it Chandra} PSF.
We find that in ObsID 11237 the nearby source contributes 14.1\% of the total flux.
We also find that this nearby source
is consistent with being constant in time
(over the {\it Chandra} epochs)
and has a mean count rate of 0.0010\,ct\,s$^{-1}$
(15\% error) in the 0.2--10\,keV band.

We speculate that this source interfered with the X-ray spectral fitting
reported by Chandra et al. (2012).
In fact, using an extraction aperture that does~not contain the nearby source
changes the result relative to an
extraction with a bigger aperture that contains the second source.
Therefore, in order to minimize the contamination, we manually selected a small
aperture ($3''$ radius) with minimal second-source 
flux (i.e., the aperture was shifted from the source center to exclude photons
from the nearby source).

\begin{figure*}
\centerline{\includegraphics[width=18cm]{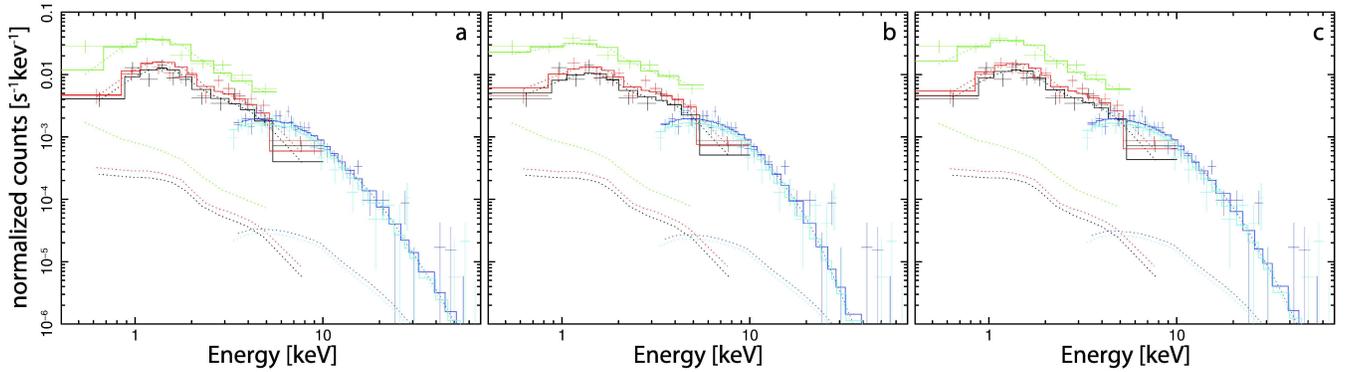}}
\caption{{\it Panel (a)}: Best-fit Mekal (zvphabs*mekal) to the combined {\it NuSTAR}$+${\it XMM} observation. The
model consists of two components: The lower dotted lines represent the
fixed power-law model of the faint nearby sources (see text), and the
upper dotted lines represent the zvphabs*mekal best-fit model to the SN\,2010jl X-ray spectrum.
The solid lines (stairs) show the best combined fit for each instrument,
while the plus signs show the data with error bars.
The instruments are {\it NuSTAR} FPM A (blue), {\it NuSTAR} FPM B
(cyan), {\it XMM} PN (green), {\it XMM} MOS1 (black), and {\it XMM} MOS2
(red). The fit parameters are listed in Table~\ref{tab:XSpec}.
{\it Panel (b)}: Like panel (a) but for zvphabs*powerlaw*spexpcut.
{\it Panel (c)}: Like panel (a) but for zvphabs*powerlaw*spexpcut with fixed $\Gamma=1$.
}
\label{fig:SN2010jl_XMM_NuStar_spec}
\end{figure*}

Table~\ref{tab:XSpec} gives a summary of our best-fit models to
the various X-ray observations. We note that
some of these models have strong degeneracies between
the parameters.
Therefore, it is hard to interpret the X-ray spectra.
Moreover, we still lack a good physical understanding of the X-ray spectra
from optically thick shocks.
Given these caveats,
in Table~\ref{tab:XSpec} we fit several models,
some of which are motivated by our modeling of the optical light curve,
presented in \S\ref{sec:ModelX}.
The {\it NuSTAR}$+${\it XMM} spectral fits are shown in
Figure~\ref{fig:SN2010jl_XMM_NuStar_spec}.
The models we use are either Mekal spectra or power laws with
an exponential cutoff which corresponds to the gas
temperature.
In addition, the models include bound-free absorption
due to solar-metallicity gas.
\begin{deluxetable*}{lllllll}
\tablecolumns{7}
\tablewidth{0pt}
\tablecaption{Spectral Modeling of the X-ray Data}
\tablehead{
\colhead{Instruments}     &
\colhead{ObsIDs}          &
\colhead{MJD}             &
\colhead{Counts}          &
\multicolumn{3}{c}{Model} \\
\colhead{}                &
\colhead{}                &
\colhead{}                &
\colhead{(cnt)}             &
\colhead{Par}             &
\colhead{Val}             &
\colhead{C-stat/dof (goodness)} 
}
\startdata
{\it Chandra}            & 11237,11122,13199 & 55536 &  485 & zphabs*zbb   &                                              & 33.0/34 (0.76) \\
                         &                   &       &      & $N_{{\rm H}}$&$20\times10^{22}$\,cm$^{-2}$ (frozen)         &                \\
                         &                   &       &      & $kT$         &$3.4_{-0.7}^{+1.2}$\,keV                      &                \\ 
                         &                   &       &      & norm         &$(4.4_{-1.5}^{+4.0})\times10^{-5}$            &                \\ \hline
                         &                   &       &      & zphabs*powerlaw*spexpcut   &                                & no fit         \\
                         &                   &       &      & $kT$         &$1.5$\,keV (frozen)                           &                \\ \hline
{\it Chandra}            & 13781             & 55852 & 1257 & zphabs*zbb   &                                              & 73.5/76 (0.20) \\
                         &                   &       &      & $N_{{\rm H}}$&$(0.7_{-0.2}^{+0.3})\times10^{22}$\,cm$^{-2}$ &                \\
                         &                   &       &      & $kT$         &$3.4_{-0.5}^{+0.7}$\,keV                      &                \\ 
                         &                   &       &      & norm         &$(4.7_{-1.3}^{+2.5})\times10^{-5}$            &                \\ \hline
                         &                   &       &      & zphabs*powerlaw*spexpcut &                                  & 81.4/76 (0.36) \\
                         &                   &       &      & $N_{{\rm H}}$&$0.99_{-0.39}^{+0.43})\times10^{22}$\,cm$^{-2}$&               \\
                         &                   &       &      & $kT$         &$15$\,keV (frozen)                            &                \\
                         &                   &       &      & $\Gamma$     &$-0.45\pm0.23$                                &                \\ 
                         &                   &       &      & norm         &$(1.41_{-0.44}^{+0.65})\times10^{-5}$         &                \\ \hline
{\it NuSTAR}$+${\it XMM} & (Table~\ref{tab:LogXray})& &     & zvphabs*mekal&                                              & 120.6/95 (0.79)\\
                         & Ignoring faint sources & &       & $N_{{\rm H}}$&$(1.1_{-0.2}^{+0.2})\times10^{22}$\,cm$^{-2}$ &                \\
                         &                   &       &      & $kT$         &$18.2_{-4.0}^{+6.2}$\,keV                     &                \\ \hline
                         & Faint sources removed &   &      & zvphabs*mekal&                                              & 119.7/94 (0.73)\\
                         &                   &       &      & $N_{{\rm H}}$&$(1.1_{-0.2}^{+0.3})\times10^{22}$\,cm$^{-2}$ &                \\
                         &                   &       &      & $kT$         &$17.7_{-3.9}^{+6.1}$\,keV                     &                \\ \hline
                         & Faint sources removed &   &      & zvphabs*powerlaw*spexpcut &                                 & 94.0/94 (0.16) \\
                         &                   &       &      & $N_{{\rm H}}$&$(0.28_{-0.17}^{+0.21})\times10^{22}$\,cm$^{-2}$&              \\
                         &                   &       &      & $kT$         &$(5.6_{-1.2}^{+1.9})$\,keV                    &                \\ 
                         &                   &       &      & $\Gamma$     &$0.45_{-0.26}^{+0.26}$                        &                \\ \hline
                         & Faint sources removed &   &      & zvphabs*powerlaw*spexpcut &                                 & 105.5/95 (0.43)\\
                         &                   &       &      & $N_{{\rm H}}$&$(0.65_{-0.14}^{+0.16})\times10^{22}$\,cm$^{-2}$&              \\
                         &                   &       &      & $kT$         &$(10.1_{-1.6}^{+2.1})$\,keV                   &                \\ 
                         &                   &       &      & $\Gamma$     &$1$ (frozen)                                  &                     

\enddata
\tablecomments{Separated by horizontal lines
are the different models fitted to the three epochs
of X-ray spectra.
Models that include redshift (e.g., zphabs, zbb)
use the SN redshift as a frozen parameter;
spexpcut is an exponential cutoff model of the form $\exp(-[E/kT]^{\gamma})$, where we freeze $\gamma=1$; and
powerlaw is a power-law model of the form $\propto E^{-\Gamma}$, where the normalization parameter
has units of photons\,keV$^{-1}$\,cm$^{-2}$\,s$^{-1}$ at 1\,keV.
Goodness is calculated using the {\tt Xspec} ``goodness 1000'' command (i.e., the fraction of realizations with C-statistic $<$ best-fit C-statistic).
The {\it NuSTAR}$+${\it XMM} fits have two versions.
Those with ``Ignoring faint sources'' in the second column are fits for all the photons within
the large extraction apertures of {\it NuSTAR} and {\it XMM}.
This fit is contaminated by the faint sources within the PSF (Fig.~\ref{fig:SN2010jl_Chandra13781}).
In fits marked by ``Faint sources removed'' we added a frozen component to the model
that takes into account the combined spectrum of all the faint sources within the PSF,
as measured in the {\it Chandra} images (see \S\ref{sec:Chandra}).
}
\label{tab:XSpec}
\end{deluxetable*}

\begin{figure}
\centerline{\includegraphics[width=8.5cm]{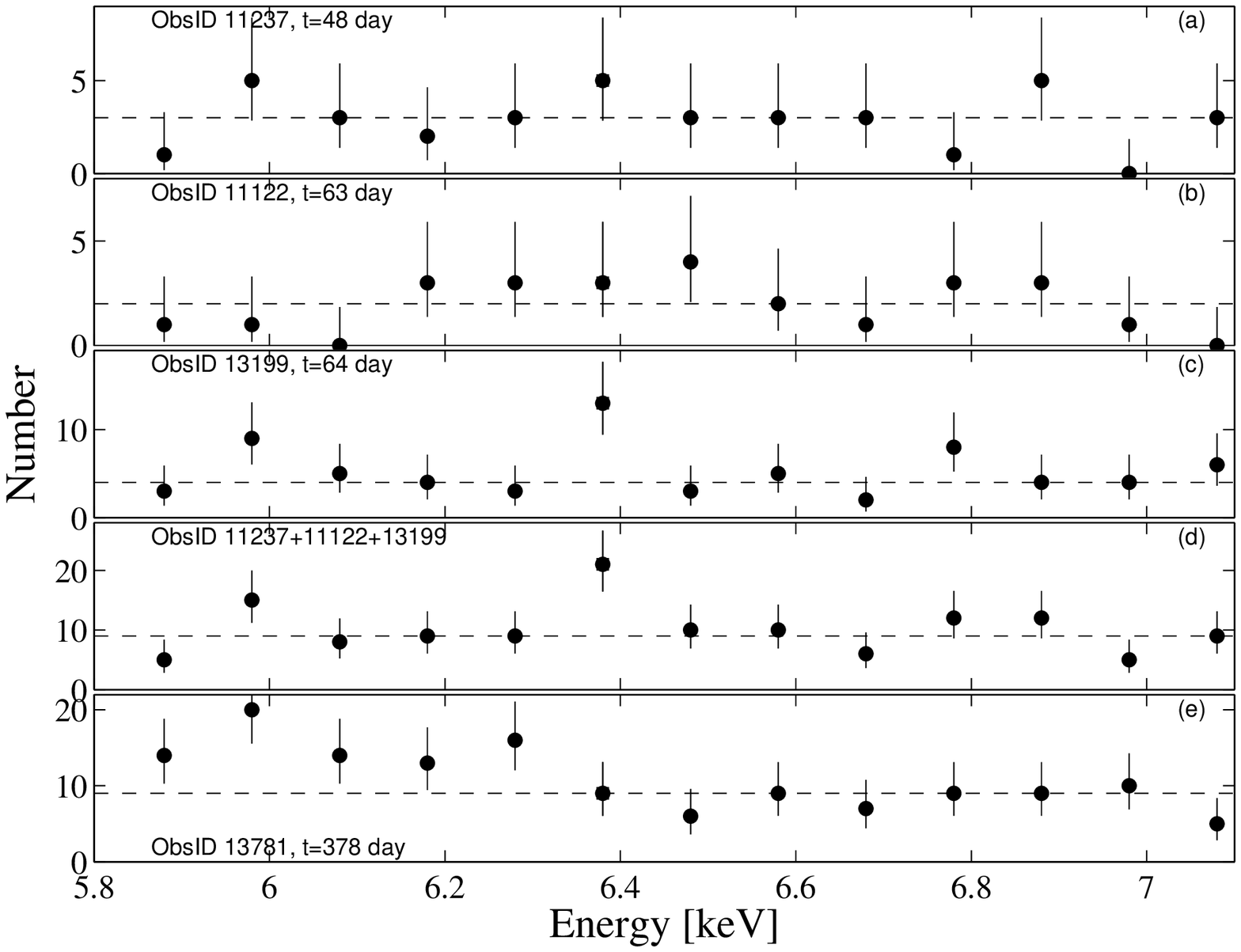}}
\caption{{\it Chandra} spectra of SN\,2010jl around the K$\alpha$-line energy.
The spectra are uncorrected for instrumental sensitivity.
The first three panels (a--c) show the individual first three observations.
Panel (d) shows the coaddition of the first three observations, designated as
the first {\it Chandra} epoch.
Panel (e) presents the second {\it Chandra} epoch.
The time relative to MJD 55474 is marked on each plot.
The bin size is 0.1\,keV which corresponds to a velocity of 4700\,km\,s$^{-1}$.
Measurements indicated by squares show the bin centered around 6.38\,keV
(i.e., the rest-frame energy of the K$\alpha$ line).}
\label{fig:Chandra_Kalpha}
\end{figure}

Chandra et al. (2012) reported a detection of a 6.38\,keV iron K$\alpha$ emission
line in the {\it Chandra} spectrum taken in the first year.
In Figure~\ref{fig:Chandra_Kalpha} we show the {\it Chandra} spectra, uncorrected for
instrumental sensitivity, around the K$\alpha$ energy.
The third observation (ObsID 13199) as well as the coaddition of the first
three observations (i.e., first epoch; ObsIDs 11237, 11122, 13199)
show a possible detection of the K$\alpha$ emission line.
In order to estimate the line properties and significance we
used the maximum-likelihood technique to fit a Gaussian profile to the line.
We find that the best-fit rest-frame energy (assuming $z=0.0107$)
is $6.41_{-0.04}^{+0.03}$\,keV,
the line width 
is $\sigma=0.033_{-0.032}^{+0.19}$\,keV (this corresponds to
$\sigma=1540_{-1500}^{+8800}$\,km\,s$^{-1}$),
and the line flux is $(3.6_{-3.0}^{+5.8})\times10^{-5}$\,counts\,s$^{-1}$.
We note that the ACIS-S energy resolution around 6.4\,keV is about 280\,eV,
which corresponds to a velocity of 13,000\,km\,s$^{-1}$.
Therefore, our best-fit line width prefers an unresolved
spectral line (i.e., zero velocity broadening).
We also find there is a 2.5\% probability that the K$\alpha$ line is not real.

\section{Model}
\label{sec:model}

Here we outline a theoretical
framework to analyze the observations
in the context of an interaction model.
We compare this model with the observations in \S\ref{sec:obsmodel}.
An important caveat for our model is that it assumes spherical symmetry,
which is reasonable only
if the deviations from spherical symmetry are of order
unity.

Our modeling strategy is similar to the one described by Svirski et al. (2012),
but it is more general in the sense that we do~not assume the values
of the CSM radial density distribution and ejecta velocity distribution.
A qualitative outline of the model is
presented in \S\ref{sec:qual}.
Section~\ref{sec:ModelOptical} presents the model quantitatively and describes the bolometric
luminosity as a function of time.
In \S\ref{sec:RadioVis} we discuss the possibility of detecting radio emission,
and \S\ref{sec:HighEnergy} discusses the properties of the X-ray emission.

\subsection{Qualitative Description of the Model}
\label{sec:qual}

A brief outline of the model is as follows.
After the SN shock moves beyond the stellar surface, it propagates in
an optically thick CSM and some of its kinetic energy is
converted into optical photons (UV to IR). 
The relevant source of opacity 
is mainly Thomson scattering, which is independent of wavelength.
If the Thomson optical depth $\tau$ is large enough, the photons
are trapped and the shock energy is mediated by photons ---
photons diffuse out, scattering upstream electrons and accelerating them.
A radiation-mediated shock ``breaks down'' or ``breaks out''
(i.e., radiation escapes ahead of the shock)
when photons diffuse ahead of the shock faster than
the shock propagates.
This happens when $\tau\approx c/v_{{\rm s}}$ (Weaver et al. 1976;
and see discussion for the case of wind-breakout in Ofek et al. 2010).
Here $v_{{\rm s}}$ is the shock velocity, and $c$ is the speed of light.

Katz et al. (2011) and Murase et al. (2011) showed that
if there is a sufficiently large amount of CSM above the shock-breakout radius,
the shock will transform from being radiation mediated
to collisionless (i.e., the photons are no longer trapped).
At this time the shock (and ejecta) is moving through the CSM and
its kinetic energy is converted to radiation
at a rate of $\epsilon(\rho v_{{\rm s}}^{2}/2)(4\pi r_{{\rm s}}^{2}v_{{\rm s}})$,
where $\epsilon$ is the efficiency,
$\rho$ is the CSM density, and $r_{{\rm s}}$ is the shock radius
(e.g., Svirski et al. 2012).
The time dependence of $r_{{\rm s}}$ and $v_{{\rm s}}$,
while the ejecta and CSM are interacting, are known
from self-similar solutions of the hydrodynamical equations (Chevalier 1982).

Later on, when the shock runs over a mass of CSM equivalent
to the ejecta mass, the shock will go into a new phase of
either conservation of energy if the density is low enough
and the gas cannot cool quickly (i.e., the Sedov-Taylor phase),
or to a conservation of momentum if the gas
can radiate its energy by fast cooling
(i.e., the snow-plow phase).
In either case, the light curve in this final stage will
be characterized by a steeper decay rate (Svirski et al. 2012).

The observables in this approach are the
light-curve rise time,
the luminosity and its decay rate,
the time of power-law break in the light curve,
and the shock velocity at late times as measured from
the hard X-ray observations.
These observables allow us to
solve for the shock radius and velocity as a function of time,
the CSM density profile, and the total mass;
they also provide a consistency test.

\subsection{The Optical Light Curve}
\label{sec:ModelOptical}

A SN explosion embedded in CSM with optical depth in excess
of $\sim c/v_{{\rm s}}$, where $c$ is the speed of light and
$v_{{\rm s}}$ is the SN shock velocity, will have
a shock breakout within the optically thick CSM.
The analytical theory behind this was presented
by Ofek et al. (2010),
Chevalier \& Irwin (2011),
Balberg \& Loeb (2011),
Ginzburg \& Balberg (2012),
Moriya \& Tominaga (2012),
and Svirski et al. (2012), while
simulations of such scenarios were presented
by Falk \& Arnett (1973, 1977), among others.
Here we review the theory and extend it to a general
CSM power-law density profile
and general ejecta velocity power-law distributions.

Following Chevalier (1982), we assume that the expanding ejecta
have a spherically symmetric power-law velocity distribution of the form
\begin{equation}
\rho_{{\rm ej}}=t^{-3}\Big(\frac{r}{tg}\Big)^{-m}.
\label{rho_ej}
\end{equation}
Here $\rho_{{\rm ej}}$ is the ejecta density,
$t$ is the time, $r$ is the radius, $m$ is the power-law index
of the velocity distribution,
and $g$ is a normalization constant.
This model is justified because the outer density profile of massive stars
can likely be approximated as a power law (e.g., Nomoto \& Sugimoto 1972).
We expect $m \approx 10$ for progenitor stars with a radiative envelope,
and $m \approx 12$ for progenitor stars with a convective envelope (e.g., Matzner \& McKee 1999).
We assume that the ejecta are expanding into a CSM with a spherically symmetric power-law
density profile of the form
\begin{equation}
\rho_{{\rm csm}} = Kr^{-w},
\label{eq:rho}
\end{equation}
where $w$ is the power-law index
and $K$ is the normalization\footnote{Chevalier (1982) denotes $K$ by $q$ and $w$ by $s$.}.
In a wind profile, $w=2$, $K=\dot{M}/(4\pi v_{{\rm CSM}})$ is called the
mass-loading parameter with units of g\,cm$^{-1}$ 
(where $v_{{\rm CSM}}$ is the CSM or wind velocity),
and $\dot{M}$ is the mass-loss rate.
We note that even if the CSM is ejected in a single outburst,
we expect the CSM to spread over a wide range of radii since
the ejecta probably have a wide range of velocities.
Given these assumptions, Chevalier (1982) showed that the
forward-shock radius is given by
\begin{eqnarray}
r_{{\rm s}} & =      & \Big(\frac{Ag^{m}}{K}\Big)^{1/(m-w)} t^{(m-3)/(m-w)} \cr
            & \equiv & r_{0} \Big(\frac{t}{t_{0}}\Big)^{(m-3)/(m-w)},
\label{eq:r_t}
\end{eqnarray}
where $A$ is a constant derived from
the self-similar solution.
The second part of the equation simply absorbs the coefficients
into arbitrary $r_{0}$ and $t_{0}$.
By differentiating Equation~\ref{eq:r_t} we get the forward-shock
velocity as a function of time,
\begin{eqnarray}
v_{{\rm s}} & =      & \frac{m-3}{m-w} \Big(\frac{Ag^{n}}{K}\Big)^{1/(m-w)} t^{(w-3)/(m-w)} \cr
            & \equiv & v_{0} \Big(\frac{t}{t_{0}}\Big)^{(w-3)/(m-w)},
\label{eq:v_t}
\end{eqnarray}
where
\begin{equation}
v_{0}\equiv \frac{m-3}{m-w} \frac{r_{0}}{t_{0}}.
\label{eq:v0}
\end{equation}

The shock breakout in a CSM environment occurs
when the Thomson optical depth is
\begin{equation}
\tau \approx c/v_{{\rm bo}},
\label{eq:tau_bo}
\end{equation}
where $v_{{\rm bo}}$ is the shock velocity at breakout
(e.g., Weaver 1976).
The expression for the Thomson optical depth, assuming $w>1$, is
\begin{equation}
\tau=\int_{r_{{\rm s}}}^{\infty}{\rho\kappa dr}=\frac{\kappa K}{w-1} r_{{\rm s}}^{1-w},
\label{eq:tau}
\end{equation}
where $r_{{\rm s}}$ is the forward-shock radius
and $\kappa$ is the opacity.
We note that for $w=2$, Balberg \& Loeb (2011) showed that
the total optical depth (taking into account the reverse-shock contribution)
is a factor of 1.55 times larger.
Chevalier (2013) argues that at late times the reverse shock may dominate the X-ray emission.
In this case the effective optical depth may be
even higher.
Effectively, this uncertainty can be absorbed into the uncertainty in the opacity
$\kappa$, which is discussed in \S\ref{sec:obsmodel}.
We note that our main conclusions do~not depend on the late-time observations.
From Equations~\ref{eq:tau_bo} and \ref{eq:tau}
we can derive an expression for $K$,
\begin{eqnarray}
K & \approx & \frac{c}{v_{{\rm bo}}\kappa}  (w-1)r_{{\rm bo}}^{w-1} \cr
  & =       & \frac{c}{\kappa}(w-1)\Big(\frac{m-w}{m-3}\Big)^{w-1}v_{{\rm bo}}^{w-2}t_{{\rm bo}}^{w-1},
\label{eq:K}
\end{eqnarray}
where the last step is obtained using Equation~\ref{eq:v0}.
Here $r_{{\rm bo}}$, $v_{{\rm bo}}$, and $t_{{\rm bo}}$
are the radius, velocity, and time scale of the shock breakout, respectively
(replacing $r_{0}$, $v_{0}$, and $t_{0}$).

The integrated CSM mass within radius $r$ or time $t$, assuming $w<3$ and star radius $r_{*}\ll r$,
is given by
\begin{eqnarray}
M  & = \int_{0}^{r}{4\pi r^{2} Kr^{-w}dr} = \frac{4\pi K}{3-w} r^{3-w} \cr
   & = \frac{4\pi K}{3-w} \Big(\frac{m-w}{m-3}\Big)^{3-w} v_{{\rm bo}}^{3-w} t_{{\rm bo}}^{(3-w)^{2}/(m-w)} t^{(3-w)(m-3)/(m-w)}. 
\label{eq:M}
\end{eqnarray}

Assuming fast cooling,
following the shock breakout the kinetic energy
is converted into radiation (bolometric luminosity)
at a rate of
\begin{equation}
L = 2\pi \epsilon r_{{\rm s}}^{2}\rho v_{{\rm s}}^{3}.
\label{eq:L}
\end{equation}
The value of the efficiency factor, $\epsilon$,
is discussed in \S\ref{sec:obsmodel}.
We note that Equation~\ref{eq:L} assumes that $v_{{\rm s}}\gg v_{{\rm CSM}}$.
Substituting the expressions for
$r_{{\rm s}}$ (Eq.~\ref{eq:r_t}),
$\rho$ (Eq.~\ref{eq:rho}),
and $v_{{\rm s}}$ (Eq.~\ref{eq:v_t}) into Equation~\ref{eq:L}
we get
\begin{equation}
L = L_{0}t^{\alpha},
\label{eq:Lt}
\end{equation}
where
\begin{equation}
\alpha\equiv \frac{(2-w)(m-3)+3(w-3)}{m-w}.
\label{eq:alpha}
\end{equation}
and
\begin{equation}
L_{0}\equiv 2\pi \epsilon K r_{{\rm bo}}^{2-w}v_{{\rm bo}}^{3}t_{{\rm bo}}^{-\alpha}.
\label{eq:L0}
\end{equation}
Using Equation~\ref{eq:v0} we can remove $r_{{\rm bo}}$ from
Equation~\ref{eq:L} and get
\begin{equation}
L_{0} = 2\pi \epsilon K \Big(\frac{m-w}{m-3}\Big)^{2-w} v_{{\rm bo}}^{5-w} t_{{\rm bo}}^{2-w-\alpha}.
\label{eq:L01}
\end{equation}
Equation~\ref{eq:Lt} was derived by Svirski et al. (2012)
for the special case of $w=2$ and $m=$12,~7,~4.

\begin{figure}
\centerline{\includegraphics[width=8.5cm]{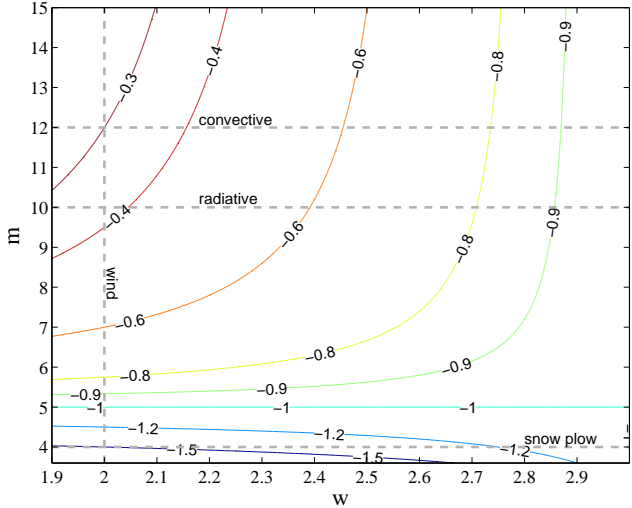}}
\caption{Contours of the value of $\alpha$ (i.e., power-law index of the early-time light curve; Eq.~\ref{eq:alpha})
as a function of $m$ and $w$.
The dashed-gray lines show several (labeled) interesting values of of $m$ and $w$.}
\label{fig:m_w_alpha}
\end{figure}

Equation~\ref{eq:Lt} provides a description of the light curve
following the shock breakout, assuming $w<3$ and $m>4$ (for radiative shock).
However, another condition is that $w\ge2$.
The reason is that if $w<2$ then the diffusion time scale diverges,
and therefore the shock will breakout near the edge of the CSM.
In this case we will not see a light curve with a power-law decay
(i.e., Eq.~\ref{eq:Lt}) lasting for a long period of times
as seen in Figure~\ref{fig:PTF10aaxf_LC}.
Therefore, $w<2$ is not a relevant solution for SN\,2010jl.
Figure~\ref{fig:m_w_alpha} presents the value of $\alpha$ as a function of $m$ and $w$.
We are not aware of a relevant self-similar
solution\footnote{The Waxman \& Shvarts (1993) solution does~not correspond to fast cooling,
which is the case here.}
for $w>3$.

Equation~\ref{eq:Lt} is correct only if the shock is in the fast-cooling regime.
The free-free cooling time scale is
\begin{equation}
t_{{\rm ff,cool}} \approx 1.8\times10^{15} \Big(\frac{T}{10^{8}\,{\rm K}}\Big)^{1/2} \Big(\frac{n}{1\,{\rm cm}^{-3}}\Big)^{-1} Z^{-2}\,{\rm s},
\label{eq:tcool} 
\end{equation}
where $Z$ is the atomic number of the atom and $n$ is the particle density
given by
\begin{equation}
n=\frac{K}{\langle \mu_{{\rm p}}\rangle m_{{\rm p}}} r^{-w},
\label{eq:n}
\end{equation}
where $\langle \mu_{{\rm p}}\rangle$ is the mean number of nucleons
per particle (mean molecular weight) and $m_{{\rm p}}$ the proton mass.
The criterion for fast cooling is that
$t_{{\rm ff,cool}}\ltorder t$.
Therefore, for time scales of a year ($3\times10^{7}$\,s),
fast cooling requires $n\gtorder6\times10^{7}$\,cm$^{-3}$.

Several other important relations can be derived.
By rearranging Equation~\ref{eq:K} we get
\begin{equation}
t_{{\rm bo}} = \Big[ \frac{c}{\kappa K} (w-1) \Big]^{1/(1-w)} \frac{m-3}{m-w} v_{{\rm bo}}^{(w-2)/(1-w)}.
\label{eq:tbo1}
\end{equation}
From Equation~\ref{eq:L01} we find
\begin{equation}
K = \frac{L_{0}}{2\pi \epsilon} \Big(\frac{m-w}{m-3}\Big)^{w-2} v_{{\rm bo}}^{w-5} t_{{\rm bo}}^{\alpha+w-2},
\label{eq:K1}
\end{equation}
and by substituting Equation~\ref{eq:K1} into Equation~\ref{eq:tbo1} we get
\begin{equation}
t_{{\rm bo}} = \Big[ 2\pi \epsilon \frac{m-w}{m-3} (w-1)\frac{c}{\kappa L_{0}} v_{{\rm bo}}^{3} \Big]^{1/(\alpha-1)},
\label{eq:tbo2}
\end{equation}
or alternatively
\begin{equation}
v_{{\rm bo}} = t_{{\rm bo}}^{(\alpha-1)/3}  \Big[ 2\pi \epsilon \frac{m-w}{m-3} (w-1)\frac{c}{\kappa L_{0}} \Big]^{-1/3}.
\label{eq:vbo2}
\end{equation}
These relations suggest that in SNe which are powered by interaction
we expect to detect correlations between the SN rise time,
its peak luminosity, and shock velocity.
We note that this can be used to test
the hypothesis that the super-luminous SN (see review in Gal-Yam 2012)
are powered by interaction of their ejecta and CSM (e.g., Quimby et al. 2011).
As far as we can tell, such correlations are not expected
in the context of other models (e.g., Kasen \& Bildsten 2010).
Furthermore, by inserting Equations~\ref{eq:K} and \ref{eq:vbo2} into Equation~\ref{eq:M},
we get the total CSM mass swept by the shock up to time $t$ as a function
of the observables (e.g., $L_{{\rm 0}}$, $t_{{\rm bo}}$),
\begin{eqnarray}
M & =      & 4\pi c \Big(\frac{m-w}{m-3}\Big)^{5/3} \frac{(w-1)^{2/3}}{3-w} \Big(\frac{2\pi c \epsilon}{L_{0}}\Big)^{-1/3} \kappa^{-2/3}\cr
  & \times & t_{{\rm bo}}^{[(m-w)(3w-4) + (w-3)(3w-6)+(2-w)(m-3)]/[3(m-w)]} \cr
  & \times & t^{(3-m)(w-3)/(m-w)}.
\label{eq:Mt1}
\end{eqnarray}
For the specific case of $w=2$ and $m=10$ we can write this as
\begin{eqnarray}
M & \approx & 15.1 \Big(\frac{\epsilon}{0.25}\Big)^{-1/3} 
                 \Big(\frac{L_{0}}{10^{46}\,{\rm erg\,s}^{-1}}\Big)^{1/3}
                 \Big(\frac{\kappa}{0.34\,{\rm cm}^{2}\,{\rm gr}^{-1}}\Big)^{-2/3}\cr
  &       &      \Big(\frac{t_{{\rm bo}}}{20\,{\rm day}}\Big)^{2/3}
                 \Big(\frac{t}{365\,{\rm day}}\Big)^{7/8}\,{\rm M}_{\odot}.
\label{eq:Mtwm}
\end{eqnarray}
We note that $L_{0}$ is the luminosity evaluated
at time of 1\,s rather than $t_{{\rm bo}}$ (see definition in Eq.~\ref{eq:Lt}).
Additional relations can be derived, including relations
that depend on $v_{{\rm bo}}$ and/or the integrated luminosity (i.e., $\int{Ldt}=L_{0}t^{\alpha+1}/[\alpha+1]$),
rather than on $L_{0}$.
However, some of these relations are algebraically long,
and we do~not provide them here.

At later times when the mass of the CSM accumulated by the ejecta is equivalent
to the ejecta mass, the light curve evolves in a different way
than described so far.
In the case of fast cooling (i.e., cooling time scale [e.g., Eq.~\ref{eq:tcool}]
is shorter than the dynamical time scale),
the system enters the snow-plow phase.
While at these times the reverse shock is absent and
the formalism of Chevalier (1982) does~not apply,
we can obtain the correct time dependence by using an artificial value of $m$
(no longer related to the ejecta profile).
In this snow-plow phase the light curve evolves effectively with $m=4$
regardless of the value of $w$ (see Svirski et al. 2012).
The reason is that, while in this case the energy is radiated away,
the momentum is conserved, and from momentum conservation
$\rho r^{3} v \approx {\rm constant}$,
we get $\rho \propto v^{-4}$, hence $m=4$.
If the shock is slowly cooling, we enter the Sedov-Taylor phase and
the light curve will drop rapidly.

Figure~\ref{fig:m_w_alpha} suggests that for $m=4$, $\alpha\approx -3/2$,
with relatively weak
dependence on the value of $w$.
However, the exact value of $\alpha$ is sensitive to the value of $m$,
and for $m$ slightly lower than $4$, $\alpha$ can change
dramatically.
In any case, once the swept-up CSM mass is comparable to the ejected mass,
we expect substantially more rapid decline of the bolometric emission.

\subsection{Visibility of a Radio Signal}
\label{sec:RadioVis}

Given the CSM density profile we can calculate some additional properties.
The column density, assuming $w>1$, between radius $r$ and infinity (i.e., the observer) is
\begin{equation}
N = \int_{r}^{\infty}{\frac{K}{\langle \mu_{{\rm p}}\rangle m_{{\rm p}}} r^{-w}dr} = \frac{K}{\langle \mu_{{\rm p}}\rangle m_{{\rm p}} (w-1)}r^{1-w}.
\label{eq:NH}
\end{equation}
The free-free optical depth between the shock region and the observer is given by (e.g., Lang 1999, Eq. 1.223; Ofek et al. 2013a)
\begin{eqnarray}
\tau_{{\rm ff}} & \approx & 8.5\times10^{-28} T_{{\rm e},4}^{-1.35} \nu_{10}^{-2.1} \int_{r}^{\infty}{n_{{\rm e}}^{2}dr} \cr
                & \cong   & 8.5\times10^{-28} T_{{\rm e},4}^{-1.35} \nu_{10}^{-2.1} \frac{K^{2}}{\langle \mu_{{\rm p}}\rangle^{2} m_{{\rm p}}^{2} (2w-1)} r^{1-2w}, 
\label{eq:tauff}
\end{eqnarray}
where $T_{{\rm e},4}$ is the electron temperature in units of $10^{4}$\,K and $\nu_{10}$ is the frequency
in units of 10\,GHz.
Note that $r$ is measured in cm, and that the last expression is valid for $w>1/2$.
If $\tau_{{\rm ff}}\gg 1$ a radio signal is not expected.

\subsection{High-Energy Emission}
\label{sec:HighEnergy}

{\it NuSTAR} opens the hard X-ray band for discovery.
Specifically, the shock temperatures associated with typical
SN shock velocities ($\sim10^{4}$\,km\,s$^{-1}$) are above 10\,keV.
Therefore, if the shock is in an optically thin region,
the X-ray temperature constitutes a reliable measurement of the shock velocity.
The shock velocity depends on the shock temperature ($kT$)
and, assuming an equation of state with $\gamma=5/3$
and an equilibrium between the electrons and protons,
is given by (e.g., Gnat \& Sternberg 2009)
\begin{eqnarray}
v_{{\rm sh}} & \approx & \sqrt{\frac{16 kT}{3 \langle\mu_{{\rm p}}\rangle m_{{\rm p}}}} \cr
           & \approx   & 2920 \Big(\frac{\langle\mu_{{\rm p}}\rangle}{0.6}\Big)^{-1/2} \Big(\frac{kT}{10\,{\rm keV}}\Big)^{1/2}\,{\rm km\,s}^{-1}.
\label{eq:vsh_kt}
\end{eqnarray}
If equilibrium between the electrons and protons is not present,
as expected in SN remnants (e.g., Itoh 1978; Draine \& Mckee 1993; Ghavamian et al. 2013),
then Equation~\ref{eq:vsh_kt}, gives a lower limit on the shock velocity.
We note that the expected equilibrium time scale between the protons
and electrons is of order
$6\times10^{8}(v_{{\rm s}}/3000\,{\rm km\,s}^{-1})^{3}n_{{\rm e}}^{-1}$\,s,
where $n_{{\rm e}}$ is the electron density in cm$^{-3}$
(i.e., roughly given by Eq.~\ref{eq:n}; Ghavamian et al. 2013).

However, if the Thomson optical depth is larger than a few, the X-ray emission
becomes more complicated.
Katz et al. (2011) and Murase et al. (2011) showed that after the shock breakout
in a wind CSM environment, the shock transforms from being radiation dominated to
collisionless, and hard X-ray emission should be generated.
However, Chevalier \& Irwin (2012) and Svirski et al. (2012) 
argued that the hard X-ray photons will be Comptonized to lower
energies, and that when the optical depth is large the X-ray spectrum
will have a cutoff above an energy of $\sim m_{{\rm e}}c^{2}/\tau^{2}$.
According to Svirski et al. (2012), the observed energy cutoff of the X-ray photons
will be
\begin{equation}
k_{{\rm B}} T_{{\rm x,obs}} \approx \min{\Big[\frac{m_{{\rm e}}c^{2}}{\tau^{2}}, \frac{3}{16}\mu_{{\rm p}} m_{{\rm p}} v_{{\rm s}}^{2} \Big]},
\label{eq:Xenergy}
\end{equation}
where the second term in the minimum function is the shock
temperature from Equation~\ref{eq:vsh_kt}.

Ignoring bound-free absorption,
Svirski et al. (2012) estimated that the X-ray luminosity is roughly given by
\begin{equation}
L_{{\rm X}}(t) \approx L(t) \frac{T_{{\rm x,obs}}(t)}{T_{{\rm e}}(t)} \min\Big[1, \frac{\epsilon^{{\rm ff}}}{\epsilon^{{\rm IC}}}(t)\Big].
\label{eq:Lx}
\end{equation}
Here $\epsilon^{{\rm ff}}$ and $\epsilon^{{\rm IC}}$ are the free-free and inverse-Compton cooling efficiencies, respectively
(see Chevalier \& Irwin 2012; Svirski et al. 2012),
and $T_{{\rm e}}$ is the electron temperature (Equation~\ref{eq:vsh_kt}).
Equation~\ref{eq:Lx} neglects the effect of bound-free absorption, and therefore should be regarded
as an upper limit.
Furthermore, we note that there is no agreement between different theoretical models
on the X-ray spectral and flux evolution.

Chevalier \& Irwin (2012)
define\footnote{The formal definition of the ionization parameter is different,
but the definition used by Chevalier \& Irwin (2012) is proportional to the ionization parameter and is used
self consistently.}
an ionization parameter as $\xi = L/(nr^{2})$.
This definition is only valid when material above the shock is optically thin.
When the optical depth (Eq.~\ref{eq:tau}) is larger than unity, one
needs to take into account the fact that the photons diffuse out slower
than the speed of light.
Since the effective outward-diffusion speed of the photons is $\sim c/\tau$,
we define the ionization parameter as
\begin{equation}
\xi \sim \frac{L}{nr^{2}} \max\Big[ \tau, 1 \Big].
\label{eq:Xi}
\end{equation}
However, we stress that this is only an order of magnitude estimate
of the ionization parameter.
Chevalier \& Irwin (2012) argue that if the ionization parameter is
larger than $\sim10^{4}$, then all the metals (which dominate the bound-free absorption)
will be completely ionized, and for $\xi\gtorder10^{2}$ the CNO elements
will be completely ionized.
Here, an important caveat is that it is not clear if the estimate of Chevalier \& Irwin (2012)
is valid for high optical depth.

\section{Modeling the Observations}
\label{sec:obsmodel}

Integrating the visible-light luminosity of SN\,2010jl
gives a lower limit on its radiated energy in the first three years
of $>9\times10^{50}$\,erg.
This is among the highest radiated bolometric energies observed for any SN
(e.g., Rest et al. 2011).
This fact, along with the long-term X-ray emission, and emission lines
seen in the optical spectra, suggest that SN\,2010jl is powered by
interaction of the SN ejecta with CSM.
Therefore, here we attempt to understand the SN observations
with the model described in \S\ref{sec:model}.

In \S\ref{sec:ModelEarly} we discuss the modeling of the first-year optical light curves;
we show 
that the model presented in \S\ref{sec:model}
describes the observations well, and that
it requires a CSM mass in excess of about 10\,M$_{\odot}$.
Section~\ref{sec:ModelLate} deals with the nature of the break in the optical light curve and the slope after the break,
and in \S\ref{sec:ModelX} we verify the consistency of the X-ray observations
with our model.

\subsection{Early Optical Light Curve}
\label{sec:ModelEarly}

In our model, the rise time is governed by
the shock-breakout time scale, and the light curve following shock breakout
is given by Equation~\ref{eq:Lt} with $m \approx 10$--12 at early times
and $m \approx 4$ at late times.
Alternatively, $m \approx 10$--12 and $w$ changes with radius.
As a reminder, we note that the value of $m$ at early times is related to
the polytropic structure of the stellar envelope
(e.g., Matzner \& McKee 1999),
while $m=4$ at late times is obtained from conservation of momentum (\S\ref{sec:ModelOptical}).

Figure~\ref{fig:PTF10aaxf_LC} suggests that the light curve
of SN\,2010jl can be described as a broken power law, with the break between 180 and 340\,days
after maximum light.
Since both Figure~\ref{fig:SN2010jl_TempRad_Spec} and Figure~\ref{fig:SN2010jl_TempRad_UVOT} suggest
that the temperature in the first year was roughly constant and close
to 9000\,K, the bolometric correction is rather
small\footnote{The bolometric correction for the PTF $R$-band magnitude is about $-0.06$, $-0.27$, and $-0.60$\,mag
for black-body temperatures of 7500, 9000, and 11000\,K, respectively.}
and constant. 
Here we adopt a constant bolometric correction of $-0.27$\,mag,
which corresponds to a black-body spectrum with $T=9000$\,K.
We apply this bolometric correction to the PTF $R$-band data
to obtain the bolometric light curve.
Later we test the stability of our solution
to this assumption.

\begin{figure}
\centerline{\includegraphics[width=8.5cm]{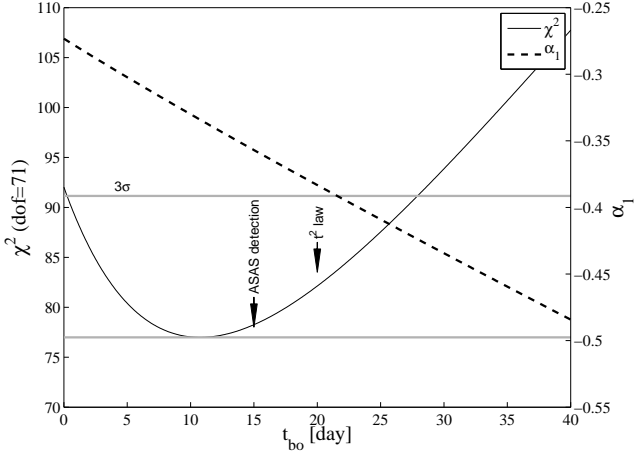}}
\caption{$\chi^{2}$ (solid line) of the fit of $L_{0,{\rm obs}} ([t-t_{{\rm bo}}]/t_{{\rm bo}})^\alpha$
as a function of $t_{{\rm bo}}$. The gray horizontal lines show the minimum $\chi^{2}$
and the $3\sigma$ confidence level assuming three free parameters.
The dashed line shows the value of $\alpha_{1}$ (the power-law index of the optical slope in the first year)
as a function of $t_{{\rm bo}}$, where its values are presented in the right-hand ordinate axis.}
\label{fig:chi2_alpha1}
\end{figure}

A power-law fit depends on the temporal zero point,
which in our model is roughly the time of maximum luminosity minus
the shock-breakout time scale.
However, since the shock-breakout time scale is related to the
light-curve rise time, and since we do~not have good constraints
on the light-curve rise time, we have to estimate the shock-breakout time
scale in a different way.
Therefore, we fitted the first-year PTF luminosity measurements with a power law 
of the form $L_{0,{\rm obs}} ([t+t_{{\rm bo}}]/t_{{\rm bo}})^\alpha$,
where $t$ is measured relative to the ASAS $I$-band maximum light (MJD 55494).
Figure~\ref{fig:chi2_alpha1} shows the fit $\chi^{2}$, as well as $\alpha_{1}$, as a function of $t_{{\rm bo}}$.
Here $\alpha_{1}$ is the power-law index of the bolometric light curve in the first year after maximum light.
The black arrows indicate $t_{{\rm bo}}$ at which the first ASAS detection was obtained,
and $t_{{\rm bo}}$ derived by fitting the first three ASAS $I$-band
measurements with a $t^{2}$ law (e.g., Nugent et al. 2011).
The fit prefers $t_{{\rm bo}}\approx10$\,day but
$t_{{\rm bo}}\ltorder25$\,day is acceptable, while the ASAS early detection indicates $t_{{\rm bo}}>15$\,day.

\begin{figure}
\centerline{\includegraphics[width=8.5cm]{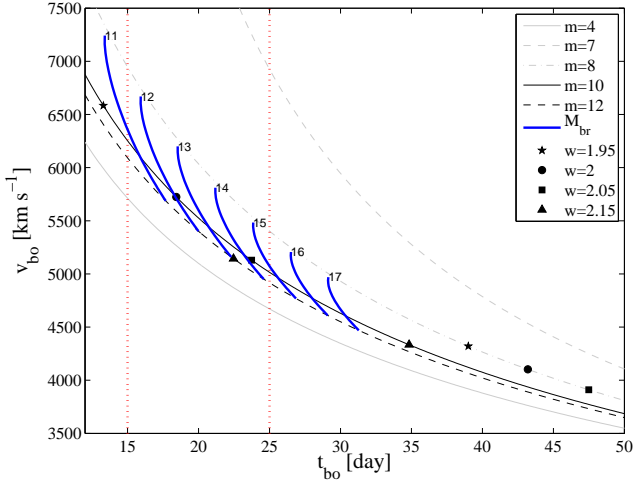}}
\caption{The solutions of Equation~\ref{eq:tbo2} as a function
of $t_{{\rm bo}}$ and $v_{{\rm bo}}$ for various values of $m$,
given the measured values of $\alpha_{1}$ (and hence $w$) and $L_{0}$ as a function
of $t_{{\rm bo}}$ (i.e., Fig.~\ref{fig:chi2_alpha1}).
Also shown, in blue contours, are lines of equal CSM mass within the break radius ($M_{{\rm br}}$),
assuming $t_{{\rm br}}=300$\,day.
The number above each contour indicates the mass in units of the solar mass.
These solutions assumes $\kappa=0.34$\,cm$^{2}$\,g$^{-1}$ and $\epsilon=1/4$.
Pentagons, circles, squares and triangles show the positions
along the various lines in which $w=1.95$, $2.0$, $2.05$ and $2.15$, respectively.
As explained in \S\ref{sec:model}, our model is valid only for $w\ge2$,
and $w<2$ can be ruled out based on the fact that the light curve has a power-law
shape with small power-law index ($\approx -0.4$) for about a year.
}
\label{fig:tbo_vbo_m}
\end{figure}

Given specific values of $\kappa$, $m$, $\alpha$, $w$, and $L_{0}$,
Equation~\ref{eq:tbo2} shows that there is a relation
between $t_{{\rm bo}}$ and $v_{{\rm bo}}$.
Moreover, based on Figure~\ref{fig:chi2_alpha1} we know that $15\ltorder t_{{\rm bo}}\ltorder25$\,day.
Figure~\ref{fig:tbo_vbo_m} shows the solutions of Equation~\ref{eq:tbo2} as a function
of $t_{{\rm bo}}$ and $v_{{\rm bo}}$ for various values of $m$,
given the measured values of $\alpha_{1}$ (and hence $w$) and $L_{0}$ as a function
of $t_{{\rm bo}}$ (i.e., Fig.~\ref{fig:chi2_alpha1}).
Also shown, in blue contours, are lines of equal CSM mass within the break radius ($M_{{\rm br}}$).
Here the break radius is defined as the radius of the shock at 300\,days --- roughly when
the observed break in the power-law light curve is detected.
Regardless of the exact values of $m$, $t_{{\rm bo}}$, and $v_{{\rm bo}}$,
Figure~\ref{fig:tbo_vbo_m} shows that the 
CSM mass $M_{{\rm br}}\gtorder10$\,M$_{\odot}$ (see also Eqs.~\ref{eq:Mt1} and \ref{eq:Mtwm}).
It also suggests that $M_{{\rm br}}\ltorder 16$\,M$_{\odot}$,
but the upper limit is somewhat weaker due to several uncertainties
that are discussed next.

Assuming $\epsilon=1/4$ and $\kappa=0.34$\,cm$^{2}$\,g$^{-1}$,
Table~\ref{tab:values} presents
the measured values of
$L_{0}$ and $\alpha$ and the calculated values of
$w$, $K$, $r_{{\rm bo}}$, $v_{{\rm bo}}$, and $M_{{\rm br}}$,
as a function of the assumed $t_{{\rm bo}}$ and $m$.
For the rest of the discussion we will adopt
$t_{{\rm bo}}=20$\,day and $m=10$.
In this case, the value of $K$ is translated to a mass-loss rate of
\begin{equation}
\dot{M}\approx 0.8 \frac{v_{{\rm CSM}}}{300\,{\rm km\,s}^{-1}}\,{\rm M}_{\odot}\,{\rm yr}^{-1},
\label{eq:Mdot}
\end{equation}
where we normalized the CSM velocity by by the highest-velocity Gaussian component in
the spectra.
This tremendous mass-loss rate is discussed in \S\ref{sec:Disc}.
\begin{deluxetable*}{lllllllll}
\tablecolumns{9}
\tablewidth{0pt}
\tablecaption{Derived SN and CSM properties}
\tablehead{
\colhead{$t_{{\rm bo}}$}     &
\colhead{$m$}                &
\colhead{$L_{0}$}            &
\colhead{$\alpha_{1}$}       &
\colhead{$w$}                &
\colhead{$K$}                &
\colhead{$r_{{\rm bo}}$}     &
\colhead{$v_{{\rm bo}}$}     &
\colhead{$M_{{\rm br}}$}     \\
\colhead{(day)}                &
\colhead{}                   &
\colhead{(erg\,s$^{-1}$)}      &
\colhead{}                   &
\colhead{}                   &
\colhead{(g\,cm$^{w-3}$)}     &
\colhead{(cm)}                 &
\colhead{(km\,s$^{-1}$)}       &
\colhead{(M$_{\odot}$)}             
}
\startdata
15  &  10 & $7.4\times10^{45}$ & $-0.36$ & \nodata & \nodata            & \nodata            &\nodata&\nodata \\
    &  12 &                    &         & 2.09    & $3.0\times10^{18}$ & $8.7\times10^{14}$ & 6100  &  9.8   \\
{\bf 20}  &  {\bf 10} & $1.2\times10^{46}$ & $-0.38$ & 2.01    & $3.0\times10^{17}$ & $1.1\times10^{15}$ & 5500  & 12.8   \\
    &  12 &                    &         & 2.13    & $1.7\times10^{19}$ & $1.0\times10^{15}$ & 5400  & 12.0   \\
25  &  10 & $1.8\times10^{46}$ & $-0.41$ & 2.06    & $2.0\times10^{18}$ & $1.2\times10^{15}$ & 5000  & 14.8   \\
    &  12 &                    &         & 2.17    & $8.9\times10^{19}$ & $1.2\times10^{15}$ & 4900  & 14.2   \\
30  &  10 & $2.7\times10^{46}$ & $-0.43$ & 2.11    & $1.2\times10^{19}$ & $1.4\times10^{15}$ & 4600  & 16.9   \\
    &  12 &                    &         & 2.21    & $4.3\times10^{20}$ & $1.3\times10^{15}$ & 4600  & 16.4   
\enddata
\tablecomments{The various parameters for different values of 
$t_{{\rm bo}}$ and $m$.
The calculations assume $\epsilon=1/4$ and $\kappa=0.34$\,cm$^{2}$\,g$^{-1}$.
The adopted values of $t_{{\rm bo}}$ and $m$ are marked in boldface.
Missing data indicate that $w<2$ and therefore the solution is not valid (see text).}
\label{tab:values}
\end{deluxetable*}

Figure~\ref{fig:tbo_vbo_m} assumes $\epsilon=1/4$.
The reason for this choice is that it is expected that a shock propagating through
the CSM will convert only the thermal energy stored in the ejecta to radiation.
The thermal energy of the ejecta is roughly half of its kinetic energy (e.g., Nakar \& Sari 2010).
In addition, since the CSM is optically thick, at early times half the photons probably diffuse
inward (and will be released at later times),
therefore taking the efficiency roughly another factor of two down.
However, at late times we expect the efficiency to increase to about $1/2$ ---
therefore, effectively $\epsilon$ may change slowly with time.
Indication for this is may be detected as as a small deviation
from the power-law decay in the first year (Fig.~\ref{fig:PTF10aaxf_LC}).
We note that the exact value of $\epsilon$ has a relatively small effect on the results.
For example, assuming $\epsilon=0.1$ ($\epsilon=0.5$) gives $M_{{\rm br}}>15$\,M$_{\odot}$ ($M_{{\rm br}}>8$\,M$_{\odot}$).

Another assumption that goes into Figure~\ref{fig:tbo_vbo_m} is that the bolometric correction
in the first year is constant.
However, as seen in Figures~\ref{fig:PTF10aaxf_LC} and \ref{fig:SN2010jl_TempRad_UVOT},
there are some indications for variations in the bolometric correction.
Given this uncertainty, we investigated the effect of variable bolometric correction on our results.
Specifically, we assumed that the effective temperature
of the photosphere evolves as $T=T_{{\rm bo}}(t/t_{{\rm bo}})^{\beta}$,
where $T_{{\rm bo}}$ is the observed temperature at shock breakout
(see \S\ref{sec:Observations}).
Assuming $T_{{\rm bo}}=9000$\,K and $t_{{\rm bo}}=20$\,day,
we corrected our light curve according to the bolometric correction we get from
the temperature, and
we investigated the effect of $\beta$ on our results.
We find that for $-0.2<\beta<0.1$ the estimate on $M_{{\rm br}}$ does~not change by more than 20\%.
Figures~\ref{fig:SN2010jl_TempRad_Spec} and \ref{fig:SN2010jl_TempRad_UVOT} suggest that $\vert \beta \vert \ltorder 0.1$.
Another unknown factor is the opacity $\kappa$.
Increasing $\kappa$ to $0.5$\,cm$^{2}$\,g$^{-1}$ will set
$M_{{\rm br}}\gtorder 8$.

The entire analysis presented here assumes that the CSM and ejecta
have spherical symmetry. This is likely not the case (e.g., Patat et al. 2011).
However, an order of unity deviation from spherical geometry will not change the results
dramatically since the integrated luminosity depends on the total mass of the CSM.
In order for the results (and specifically the $M_{{\rm br}}$ estimate)
to change significantly, an extreme geometry is probably required.
We cannot rule out such a scenario.
However, given that our model explains the observed broken-power-law
behavior, finds values of $m$ and $w$
which are consistent with expectations,
and successfully predicts the observed shock velocity
(see also \S\ref{sec:ModelLate} and \S\ref{sec:ModelX}),
we conclude that our description is correct.
Another important point may be the clumpiness of the CSM.
However, if the Chevalier (1982) solutions are still valid
on average, our results are correct, as they depend on
the global (average) properties of the CSM and ejecta.
Therefore, we conclude that our main result that the mass in the CSM
of SN\,2010jl is in excess of about 10\,M$_{\odot}$ is robust.
Finally, we note that Svirski et al. (2012) predict that
at early times the color temperature will evolve slowly with time.
This is roughly consistent with the observations of SN\,2010jl.

\subsection{Late-Time Light Curve}
\label{sec:ModelLate}

Around 300\,days after maximum light, the optical light curve of SN\,2010jl
shows a break in its power-law evolution, and the $R$-band
power-law index becomes $\alpha_{2} \approx -3$. 
The change in power-law slope at late times may have three possible explanations:
(i) we reached the snow-plow phase, and therefore $m$ changes to about $4$ (Svirski et al. 2012);
(ii) the shock became slow cooling and therefore the light curve drops rapidly;
and (iii) the shock reached the end of the CSM, or in other words, the CSM density profile
became steeper than $r^{-2}$.
Next, we will test these possibilities and find that the snow-plow phase
option is the most likely.
We note that the measurement of $M_{{\rm br}}$ is not affected by
the nature of the break.

\begin{figure}
\centerline{\includegraphics[width=8.5cm]{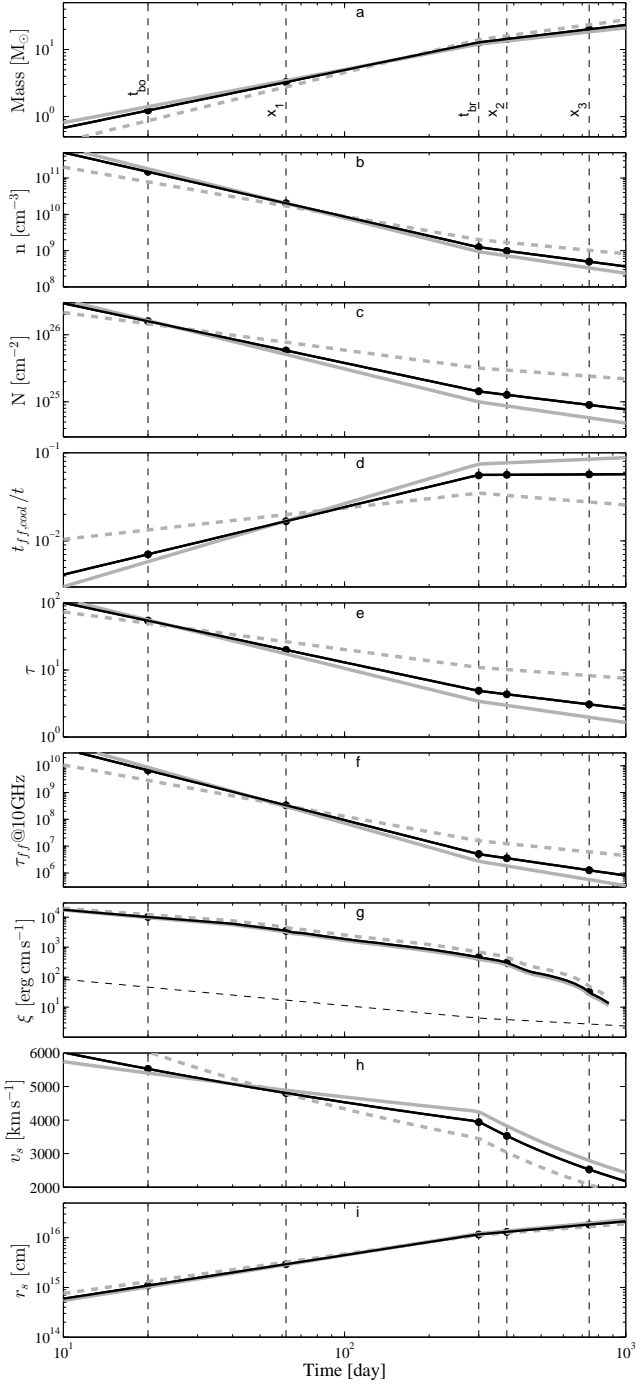}}
\caption{The CSM properties as a function of time,
assuming $t_{{\rm bo}}=20$\,day and $m=10$
(black lines).
The gray lines are for $m=12$, while the dashed-gray line is for $m=8$.
The black-dashed vertical lines show the breakout time scale ($t_{{\rm bo}}$),
the time of the optical light-curve break (300\,days; $t_{{\rm br}}$),
the times of the first two {\it Chandra} epochs ($x_{1}$ and $x_{2}$),
and the {\it XMM}$+${\it NuSTAR} epoch ($x_{3}$). The different panels
show the following:
(a) CSM mass within the shock radius,
(b) density of the CSM at the shock radius, 
(c) column density between the shock and the observer,
(d) free-free cooling time scale divided by the time at the shock radius,
(e) Thomson optical depth between the shock radius and the observer,
(f) 10\,GHz free-free optical depth,
(g) ionization parameter (Eq.~\ref{eq:Xi}),
(h) shock velocity, and 
(i) shock-radius evolution.
Time is measured relative to maximum $I$-band light minus $t_{{\rm bo}}$.
We note that panel (g) shows an additional dashed black line; it represents
the minimal ionization parameter (for $m=10$) as estimated by replacing the luminosity
in Equation~\ref{eq:Xi} by the observed X-ray luminosity ($L_{{\rm X}}\approx1.5\times10^{41}$\,erg\,s$^{-1}$).
The intrinsic X-ray luminosity may be much higher because of, for example, bound-free absorption.}
\label{fig:CSM_prop_r}
\end{figure}

Our solution suggests that the CSM density at $r_{{\rm br}}$
is $\sim10^{9}$\,cm$^{-3}$.
Given this very high density, the shock must be fast cooling and option (ii)
can be ruled out (Eq.~\ref{eq:tcool}; see also Fig.~\ref{fig:CSM_prop_r}, panel d).
Assuming $m=10$, Equation~\ref{eq:alpha} suggests that in order to get
the observed value $\alpha_{2}\approx -3$, we require $w\approx 5$.
However, the Chevalier (1982) self-similar solutions are invalid for $w>3$.
Nevertheless, the steep value of $\alpha_{2}$ probably means that if $m\approx10$, $w>3$.
We note that in this case, the shock will accelerate,
and at late times we expect
$v_{{\rm s}}\gtorder 4000$\,km\,s$^{-1}$ (Fig.~\ref{fig:CSM_prop_r}).
This is somewhat higher than the velocity suggested by our
{\it NuSTAR} observations (see \S\ref{sec:ModelX}).

Given the solution presented in Figure~\ref{fig:tbo_vbo_m} (using $\alpha_{1}$),
integration of the mass to the break radius gives $M_{{\rm br}}\gtorder 10$\,M$_{\odot}$.
Normal SN explosions have an ejecta mass that is similar, to an order of magnitude,
to our derived CSM mass.
Therefore, it is likely that the
ejecta collected a CSM mass which is equivalent to its own mass
and the system reached the snow-plow phase, hence
there is a natural explanation to the change in $\alpha$
without changing $w$ (at least not in a major way).
Of course, it is possible that during $t_{{\rm br}}$ the values of both $m$ and $w$ are changing.
This idea requires a coincidence between two independent phases and therefore
we will not discuss it further.

Assuming $w=2$ and $m=4$, we expect $\alpha_{2}\approx -3/2$ (see also Svirski et al. 2012),
while we observed $\alpha_{2}\approx-3$.
There are several possibilities to explain this.
First, at late times (a year after peak brightness) there may be significant evolution
in the bolometric correction.
Interestingly, the late-time spectra (see Fig.~\ref{fig:SN2010jl_Spec}) 
suggest that the SN becomes bluer at late times.
We note, however, that these late-time measurements are affected by
the underlying star-forming region and are therefore uncertain.
In addition, the missing radiation may be emitted in the X-ray band.
We find that if the intrinsic unabsorbed X-ray luminosity
of the SN is $\sim20$ times higher than observed,
the contribution of the X-ray luminosity to the bolometric
light curve will modify $\alpha_{2}$ to about $-3/2$.

A second possibility is that the system is approaching the slow cooling
stage and some of the energy is not released efficiently as optical photons.
Our estimate
suggests that at late times the cooling time scale is increasing to about 10\%
of the dynamical time scale (Fig.~\ref{fig:CSM_prop_r}).
Therefore, it is possible that the shock starts to be nonradiative,
hence explaining the steeper than expected power-law slope.
To summarize the issue, we suggest that the most likely explanation to the
discrepancy between the observed and predicted value of $\alpha_{2}$, is that at late times
there is a substantial bolometric correction, and possibly the shock is becoming nonradiative.
Unfortunately, we do~not have reliable multi-band or spectroscopic observations during the second year.

Based on our simple model, Figure~\ref{fig:CSM_prop_r} shows the evolution of the various
parameters as a function of time.
Panel (b) indicates that even at late times, about three years after maximum light,
the density of the CSM at the shock radius is of order a few times $10^{8}$\,cm$^{-3}$.
Interestingly, the Thomson optical depth above the shock, three years after maximum light,
is decreased to roughly unity.
This may explain why the visible-light spectrum of the SN is becoming bluer,
as the region heated by the shock is becoming more exposed and the photons
emitted in the shock region are affected by less and less processing.
The free-free optical depth [panel (f)]
above the shock at 10\,GHz, three years after maximum light,
is $\tau_{{\rm ff}} \approx 10^{5}$, assuming the electron temperature above the shock is $10^{4}$\,K.
Therefore, naively, radio emission is not expected in the near future.
However, if the electron temperature just above the shock is significantly
higher and the CSM cocoon is terminated at a few times
the shock radius, then $\tau_{{\rm ff}}$ can be small enough and radio emission would be detected.
Finally, we note that the cooling time scale divided by the hydrodynamical time scale [panel (d)]
suggests that at late times, the system may approach slow cooling, so some energy losses
(not in optical radiation) are expected.

An interesting point to note is that Figures~\ref{fig:SN2010jl_TempRad_Spec}
and \ref{fig:SN2010jl_TempRad_UVOT}
show that at late times
the effective black-body radius is decreasing.
Svirski et al. (2012) argue that at late times the fraction of the energy
released in X-rays is increasing (as seen in SN\,2010jl).
In this case, the optical photons will deviate from a black-body spectrum
as fewer photons are available in the optical,
and this can generate an apparent decrease in the effective black-body radius.
In general, this effect should caution against the use of black-body fits
to estimate the photospheric radius of such explosions.

\subsection{Modeling the X-ray Data}
\label{sec:ModelX}

We still do~not have a good theoretical understanding of the expected X-ray spectral evolution from
optically thick sources
(e.g., Katz et al. 2011; Chevalier \& Irwin 2012; Svirski et al. 2012).
%
Another problem is that the X-ray spectral observations are hard to
model. The reasons are the low number of photons, contamination
from nearby sources, and the degeneracy between the free parameters in the
various models.
Nevertheless, it is interesting to compare the rough expectations
with the observations.
Given these issues, our approach is to use the
model we constructed based on the optical data to make some
predictions for the X-ray band, and to compare the X-ray observations with these
predictions.
Especially interesting are the {\it NuSTAR}$+${\it XMM}
observations which cover a large energy range
and were taken when the Thomson optical depth
is expected to be relatively low, $\tau \approx 3$.
Here we discuss the bound-free absorption,
the X-ray flux, and the X-ray spectrum.

Figure~\ref{fig:CSM_prop_r} shows that the
predicted column density above the shock is very large,
$\sim10^{26}$\,cm$^{-2}$ during the shock breakout,
dropping to $\sim10^{25}$\,cm$^{-2}$ during our
{\it XMM}$+${\it NuSTAR} observations.
These predicted column densities are larger,
by about two orders of magnitude, than the
bound-free column densities suggested by Chandra et al. (2012) at early times.
A plausible explanation is that the CSM above the shock
is ionized by the SN radiation field.
Indeed, panel (g) in Figure~\ref{fig:CSM_prop_r} suggests
that at early times the ionization parameter (Eq.~\ref{eq:Xi}) is
$>10^{2}$\,erg\,cm\,s$^{-1}$, and possibly as high as
$\sim10^{4}$\,erg\,cm\,s$^{-1}$.
Such a large value is enough to ionize all the metals in the CSM
(Chevalier \& Irwin 2012).
However, at late times, the ionization parameter is
only $\sim10^{2}$\,erg\,cm\,s$^{-1}$, which may leave
some bound electrons in heavy elements.

The next simple test is to use the order of magnitude estimate in
Equation~\ref{eq:Lx} to predict
the X-ray flux as a function of time.
The prediction is shown in Figure~\ref{fig:SN2010jl_XLC_MeanE}
as a gray line.
At early times, about 100\,days after the SN maximum visible light,
the prediction is consistent with the observations.
About a year later, the X-ray prediction is a factor of four higher
than the observations, while around 2.5\,yr after maximum visible light,
the predicted X-ray luminosity is a factor of two higher than observed.
We note that Equation~\ref{eq:Lx} is an order of magnitude
estimate of the luminosity in the entire X-ray band [including soft and hard ($>10$\,keV) X-rays],
and that it does~not take into account the bound-free absorption,
which even if not very high, still can affect the emission
of soft X-rays considerably.
For example, for $N_{{\rm H}}=10^{22}$\,cm$^{-2}$,
the bound-free optical depth (e.g., Morrison \& McCammon 1983)
at 0.5\,keV (1\,keV) is 7.3 (2.4), which will
decrease the observed X-rays at this energy by a factor of 1600 (11).

\begin{figure}
\centerline{\includegraphics[width=8.5cm]{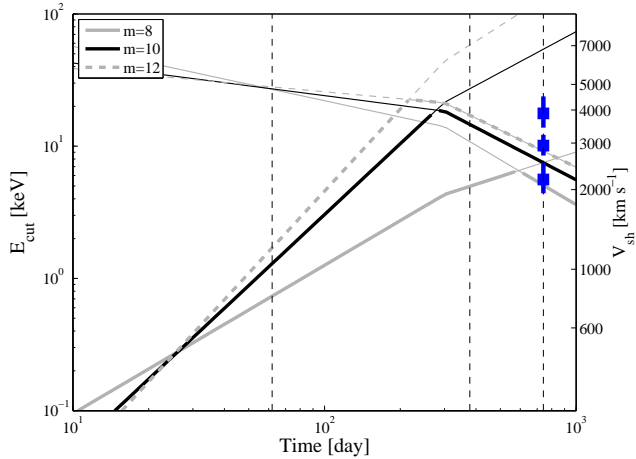}}
\caption{The predicted X-ray cutoff energy (Eq.~\ref{eq:Xenergy}) as a function of time.
Different line types and gray-scale levels are for different values of
$m$ as indicated in the legend.
The heavy lines represent the minimum function (Eq.~\ref{eq:Xenergy}),
while the thin lines represent the two possibilities in Equation~\ref{eq:Xenergy}
without taking the minimum.
The blue squares show the temperature as measured in the {\it NuSTAR}$+${\it XMM}
epoch (Table~\ref{tab:XSpec}; faint sources removed).
The upper square is for the zvphabs*mekal model,
the bottom square refers to zvphabs*powerlaw*spexpcut,
and the square in the middle is for zvphabs*powerlaw*spexpcut
with $\Gamma=1$.
The vertical dashed lines show the epochs of the two {\it Chandra}
and the {\it NuSTAR}$+${\it XMM} observations.
The right-hand ordinate axis gives the shock velocity corresponding to the cutoff energy,
based on Equation~\ref{eq:vsh_kt}.}
\label{fig:T_Ecut}
\end{figure}

According to Svirski et al. (2012),
at early times we expect that the cutoff energy will be around
$m_{{\rm e}}c^{2}/\tau^{2}$, while when the optical depth decreases
to roughly a few, we expect that the cutoff energy
will represent the shock temperature (Eq.~\ref{eq:Xenergy}).
Figure~\ref{fig:T_Ecut} shows the predicted cutoff energy as a function
of time.
Also plotted are the {\it NuSTAR}$+${\it XMM}
measured X-ray temperatures based on the various fits (Table~\ref{tab:XSpec})
and assuming temperature equilibrium between the ions and the electrons.
If equilibrium is not present, then our measurement is only
a lower limit on the shock velocity.

Figure~\ref{fig:T_Ecut} suggests that the {\it NuSTAR}$+${\it XMM}
observation measures the shock temperature, and hence the shock velocity.
The three models in Table~\ref{tab:XSpec} in which the faint nearby sources
were removed suggest a shock velocity with 1$\sigma$ confidence interval in the range
1900--4500\,km\,s$^{-1}$.
We suggest that the most physically motivated model is the power-law model
with exponential cutoff, in which the power-law index is set to $\Gamma=1$.
The reason is that below the cutoff energy we speculate that
free-free processes, with a spectrum $n(E)\propto E^{-1}$,
will dominate the emission (Svirski et al. 2012).
This model suggest an exponential cutoff energy $kT=10.1_{-1.6}^{+2.1}$\,keV,
which translates to $v_{{\rm s}}\approx2900\pm300$\,km\,s$^{-1}$.
However, if the ions and the electrons are not in equilibrium,
all we can say is that the shock velocity is larger than $\sim 2000$\,km\,s$^{-1}$. 
This measured shock velocity is in agreement with the predicted
shock velocity of $\sim 2600$\,km\,s$^{-1}$ (Figure~\ref{fig:CSM_prop_r}, panel h).
Under the assumption that the SN is powered by interaction,
by comparing the kinetic energy to the integrated luminosity,
the X-ray-derived velocity along with the integrated bolometric luminosity,
can be used to roughly determine the CSM mass.
While lacking the exact prefactors we derived in \S\ref{sec:ModelOptical},
we obtain an order of magnitude estimate of the CSM mass --- $\sim 10$\,M$_{\odot}$.

We estimate that during the {\it NuSTAR}$+${\it XMM}
observation the ionization parameter was $\sim10^{2}$\,erg\,cm\,s$^{-1}$.
According to Chevalier \& Irwin (2012) this value is not enough
to ionize all of the metals.
Therefore, our estimate of the ionization parameter is in conflict
with the value of the bound-free column density we deduced
from the {\it NuSTAR}$+${\it XMM} observations.
Possible solutions include the existence of even harder photons in the shock,
or that the estimate of the effective ionization parameter
at high optical depth is wrong.

Given the difficulties in modeling the early-time data
obtained by {\it Chandra},
we attempt to fit these observations
with a power law having an exponential cutoff
as predicted by Equation~\ref{eq:Xenergy}
(Fig.~\ref{fig:T_Ecut}) --- 1.5\,keV and 15\,keV,
for the {\it Chandra} first and second epochs, respectively.
While the fit to the second epoch has an acceptable
$C$-statistic (see Table~\ref{tab:XSpec}),
fitting the first epoch while freezing the cutoff energy at 1.5\,keV failed.
Given the unknowns associated with the X-ray emission at such
high optical depth ($\tau \approx 20$),
we do~not consider this to be a problem for our model.

We note that the marginal detection of
the K$\alpha$ line in the first {\it Chandra} epoch (\S\ref{sec:xspec}) is not naturally
explained in our model.
In the context of our model, the K$\alpha$ line must be generated
at relatively large radii where the optical depth is low.

\subsection{Emission-Line Spectra and Precursor}
\label{sec:EmLines}

The spectra of SN\,2010jl show a variety of emission lines.
Based on spectropolarimetric observations, Patat et al. (2011)
suggested that the Balmer lines form above the photosphere.
Therefore, the emission from the Balmer lines will not constitute
a good estimate of the mass in the CSM (see discussion in Ofek et al. 2013c).
Smith et al. (2012) show that the line shape evolves with time, presumably
due to the formation of dust.

Nevertheless, the width of the Balmer lines gives us an estimate of the
CSM velocity.
This is important in order to estimate when the CSM was ejected
from the SN progenitor.
Given the velocities of the Balmer lines of SN\,2010jl
between $\sim 70$\,km\,s$^{-1}$ and $\sim300$\,km\,s$^{-1}$; \S\ref{sec:spec}),
and the typical radii of the CSM of $\sim2\times10^{16}$\,cm,
we estimate that the CSM was ejected from the progenitor
$\sim 10$--100\,yr prior to the explosion.
Given this prediction, we searched for archival images at this sky location.
PTF images of the SN location taken about 200\,days prior to explosion
did~not reveal any pre-explosion outburst; see Ofek et al. (in prep.) for details.

\section{Summary and Discussion}
\label{sec:Disc}

We present optical and X-ray observations of SN\,2010jl (PTF\,10aaxf).
We extend the model described by Svirski et al. (2012) for
a SN shock interacting with an optically thick CSM.
Our model treats many of the unknowns in the problem as
free parameters.
We show that this model explains many of the details
in the optical and X-ray data.
Most interestingly, using this model we find that the mass
in the CSM must be larger than $\sim 10$\,M$_{\odot}$,
and possibly smaller than 20\,M$_{\odot}$.
This large amount of mass must have been ejected from the SN progenitor
several decades prior to its explosion.
We note that preliminary results based on the radiation hydrodynamics light-curve 
code described by Frey et al. (2013) support our results regarding the
large CSM mass required to power SN\,2010jl (Even et al., in prep.)

Our model demonstrates that the optical light curves of SNe~IIn
driven by interaction of the SN ejecta with optically thick CSM
are characterized by long-lived power laws.
Furthermore, the optical light curves
can be used in a straightforward way to measure the properties
of the CSM as well as the SN shock velocity and its evolution with time.
We note that the shock velocity is directly related to the energetics of the explosion.
We argue that measurements of the shock velocity based on spectral line widths
are likely not as accurate as this method, since they depend on where the spectral lines
are forming.

SN\,2010jl is the first SN to be detected in the hard X-ray band using {\it NuSTAR}.
The {\it NuSTAR} observation combined with {\it XMM} data taken
roughly at the same time enable us to measure the temperature of this emission.
From our model, we show that this temperature likely represents the shock velocity,
and that the measured shock velocity of $\sim3000$\,km\,s$^{-1}$ is consistent
with the prediction of our model, based on the optical data alone.
This demonstrates the power of hard X-ray observations to measure the SN shock velocity,
and possibly even the evolution of the shock velocity with time.

Interestingly, our modeling prefers solutions with CSM density profiles
$\propto r^{-2}$ (i.e., wind-like profile).
This means that either the CSM was ejected in a continuous process,
or multiple bursts,
or in a concentrated burst with a {\it velocity} distribution having a power-law index
of $\sim2$, and in which the ratio
between the velocity of the fast and slowly moving ejecta is at
least a factor of 20.
This factor is required in order to explain the shock emission which
was probed from a distance of $\sim 10^{15}$\,cm
up to more than $\sim 2\times10^{16}$\,cm.

Several mechanisms have been suggested to explain
the presence of large amounts of CSM
around SN progenitors.
Quataert \& Shiode (2012) propose that dissipation of
gravity waves originating from the stellar core can unbind
large amounts of mass.
Chevalier (2012) suggest that a common-envelope phase just prior
to explosion may be responsible for the CSM.
Soker \& Kashi (2013) argue for outbursts driven by binary star
periastron passages, and Arnett \& Meakin (2011) show that shell oxygen burning in
massive stars gives rise to large fluctuations in the turbulent
kinetic energy that in turn may produce bursts.
The most thoroughly explored mechanism is probably
the pulsational pair instability
(Rakavy, Shaviv, \& Zinamon 1967;
Woosley, Blinnikov, \& Heger 2007;
Waldman 2008), which
predicts that some massive stars will eject material several times
before their final and last explosion.
Given the large amount of CSM involved, it is possible that
SN\,2010jl is a result of multiple pulsational pair instabilities
taking place over the past several decades.
Multiple mass ejection events are required in order to explain
the average $r^{-2}$ CSM radial distribution over a factor of 20
in radii.
However, other explanations may exist (e.g., Quataert \& Shiode 2012;
Chevalier 2012).

\acknowledgments

E.O.O. thanks Roni Waldman, Nir Sapir, and Orly Gnat for discussions.
This work was supported under NASA Contract No. NNG08FD60C, and made use of data from the NuSTAR mission, a project led by the California Institute of Technology, managed by the Jet Propulsion Laboratory, and funded by NASA.
We thank the {\it NuSTAR} Operations, Software, and Calibration teams for support with the execution and analysis of these observations.
This research has made use of the {\it NuSTAR} Data Analysis Software (NuSTARDAS) jointly developed by the ASI Science Data Center (ASDC, Italy) and the California Institute of Technology (USA).
This paper is based on observations obtained with the
Samuel Oschin Telescope as part of the Palomar Transient Factory
project, a scientific collaboration between the
California Institute of Technology,
Columbia University,
Las Cumbres Observatory,
the Lawrence Berkeley National Laboratory,
the National Energy Research Scientific Computing Center,
the University of Oxford, and the Weizmann Institute of Science.
Some of the data presented herein were obtained at the W. M. Keck
Observatory, which is operated as a scientific partnership among the
California Institute of Technology, the University of California,
and NASA; the Observatory was made possible by the generous
financial support of the W. M. Keck Foundation.  We are grateful for
excellent staff assistance at Palomar, Lick, and Keck Observatories.
E.O.O. is incumbent of
the Arye Dissentshik career development chair and
is grateful to support by
a grant from the Israeli Ministry of Science and
the I-CORE Program of the Planning
and Budgeting Committee and The Israel Science Foundation (grant No 1829/12).
A.V.F.’s SN group at UC Berkeley has received generous financial
assistance from Gary and Cynthia Bengier, the Christopher R. Redlich
Fund, the Richard and Rhoda Goldman Fund, the TABASGO Foundation,
and NSF grant AST-1211916.

\end{document}